\documentclass[a4paper,11pt]{article}
\pdfoutput=1 

\usepackage{jheppub} 

\usepackage[T1]{fontenc}
\usepackage{caption}
\usepackage{subcaption}
\usepackage{float, color, colortbl}
\definecolor{Gray}{gray}{0.9}
\usepackage{bm}
\usepackage{nicefrac}
\usepackage{multirow}

\newcommand{\beq}{\begin{equation}}
\newcommand{\eeq}{\end{equation}}

\newcommand{\bea}{\begin{eqnarray}}
\newcommand{\eea}{\end{eqnarray}}

\usepackage{amsmath}

\usepackage[normalem]{ulem}
\usepackage{slashed}

\subheader{\normalsize
	\vspace*{-3.7em}
	\begin{flushright}
		IRMP-CP3-22-55\\
	\end{flushright}
}

\title{\boldmath Renormalisation group effects on SMEFT interpretations of LHC data}

\author[a]{Rafael Aoude,} 
\author[a,b]{Fabio Maltoni,}
\author[a]{Olivier Mattelaer,}
\author[c]{Claudio Severi,}
\author[c]{Eleni Vryonidou}
\affiliation[a]{Centre for Cosmology, Particle Physics and Phenomenology (CP3),\\
Universit\'e Catholique de Louvain, B-1348 Louvain-la-Neuve, Belgium}
\affiliation[b]{Dipartimento di Fisica e Astronomia, Universit\`a di Bologna and INFN,\\ 
Sezione di Bologna, via Irnerio 46, 40126 Bologna, Italy}
\affiliation[c]{Department of Physics and Astronomy, University of Manchester, \\
Oxford Road, Manchester M13 9PL, United Kingdom}
\emailAdd{rafael.aoude@uclouvain.be}
\emailAdd{fabio.maltoni@uclouvain.be}
\emailAdd{olivier.mattelaer@uclouvain.be}
\emailAdd{claudio.severi@manchester.ac.uk}
\emailAdd{eleni.vryonidou@manchester.ac.uk}

\abstract{
We explore the impact of Renormalisation Group (RG) effects in the Standard Model Effective Field Theory (SMEFT) interpretations of LHC measurements. We  implement the RG running and mixing for the Wilson coefficients as obtained from the one-loop anomalous dimension matrix in the SMEFT into a Monte Carlo generator. This allows to consistently  predict  and combine in global fits collider observables characterised by different scales. As a showcase, we  examine the impact of RG running in the strong coupling on the SMEFT predictions for $t \bar t$ production cross sections and differential distributions as well as on the bounds  on the Wilson coefficients that can be obtained from current LHC data.
}
\begin{document} 

\maketitle

\clearpage

\section{Introduction}

In the light of no evidence for the existence of  new degrees of freedom at the weak scale or below, the Standard Model Effective Field Theory (SMEFT) \cite{Weinberg:1979sa,Buchmuller:1985jz,Grzadkowski:2010es} provides a conceptually simple, compelling, and powerful framework to probe beyond the Standard Model physics. The SMEFT allows us to consistently and systematically parameterise possible deviations from the SM predictions in the interactions among the known particles, using minimal theoretical assumptions.   

The interest in the SMEFT approach has triggered considerable efforts over the last years not only in the quest for the best SM predictions, which are necessary to detect deviations, but also to improve the accuracy of the SMEFT predictions by consistently including higher order corrections in QCD and EW couplings. 
 
With improved predictions at hand and more and more precise measurements coming from the LHC,  performing global interpretations of LHC measurements has become imperative and first results in the top quark sector \cite{Buckley:2016cfg,Buckley:2015lku,Hartland:2019bjb,Brivio:2019ius}, the Higgs and electroweak gauge sector \cite{Biekoetter:2018ypq,Ellis:2018gqa,daSilvaAlmeida:2018iqo} as well as combinations of the two \cite{Ellis:2020unq,Ethier:2021bye} have appeared. 

These first studies have demonstrated that global interpretations in the SMEFT are feasible and that order of tens of operator coefficients can be determined simultaneously, by also exploiting the crucial fact that the SMEFT correlates observables from different sectors. 

 Whilst the combination of different processes is needed to maximise the potential of SMEFT interpretations, it comes with various complications and challenges. One such complication is the fact that observables are typically associated to specific energy scales, even in the same experiment. The same SMEFT operators  are therefore probed at different scales.  In order to consistently combine the results, however, Renormalization Group (RG) effects should be taken into account, as RG Equations (RGE) are necessary to account for different natural scales of different processes.

In principle, an approximate RGE flow can be computed off-line and once and for all. The complete RGE of the SMEFT at dimension-6 are known at one-loop \cite{Jenkins:2013zja,Jenkins:2013wua,Alonso:2013hga}, and several codes exist which allow one to input a set of Wilson coefficients at a given scale and extract them at a different one \cite{Celis:2017hod, Fuentes-Martin:2020zaz,Aebischer:2018bkb,Lyonnet:2013dna,DiNoi:2022ejg}.
Results on the SMEFT RGE for selected dimension-8 operators are also available at one loop \cite{Chala:2021pll,AccettulliHuber:2021uoa, DasBakshi:2022mwk, Helset:2022pde}.
Up to now, the RGE evolution in SMEFT interpretations of LHC measurements has either been neglected altogether, or it has been taken from a high-scale $\mu_0$ to a {\it fixed} low-scale $\mu$. Analyses where the scale $\mu$ is chosen bin-by-bin for differential distributions have started to appear in the literature \cite{Battaglia:2021nys}. 
However, the analysis of observables such as differential distributions, that span orders of magnitude in energy, calls for an event-by-event choice of renormalization scale, that can only be handled in a Monte Carlo tool at runtime. A dynamical scale choice requires recomputing the Wilson coefficients at every phase-space point, and the only practical way of incorporating such RGE effects into theoretical predictions is to include them into suitable MC generators. Up to now, no dedicated implementation has been made available. 

In this work we present the first implementation of RGE effects in a Monte Carlo generator, \texttt{Madgraph5\_aMC@NLO} \cite{Alwall:2014hca}. We discuss the implementation and then present phenomenological examples where the impact of RGE is investigated within the context of SMEFT interpretations, and compare it with the next-to-leading order predictions. RG improved predictions for SMEFT can potentially form an intermediate step towards a full next-to-leading (NLO) order computation by resumming large logarithms. Our current implementation focuses on leading-order RGE improved results, which capture the leading effects arising from the presence of separated energy scales. We nevertheless envision that our setup, combined with the NLO computations of  \cite{Degrande:2020evl} and the two-loop anomalous dimension matrix (once available) will form the basis of state-of-the-art SMEFT predictions in the coming years. 

The paper is organised as follows. We describe the setup used and implementation details in Sec. \ref{Sec:Setup}. In Sec. \ref{Sec:RGE_top} we discuss RGE effects for top quark operators presenting the relevant anomalous dimension matrix and several examples of operator running and mixing. In Sec. \ref{Sec:Results_LHC}, as an example we focus on top pair production and show the results for the LHC taking into account running and mixing for different choices of dynamical and fixed scales. These results are then used in Sec. \ref{Sec:ToyFit} to perform a toy fit to illustrate the impact of RGE effects when constraining the Wilson coefficients. Finally we conclude in Sec. \ref{Sec:Conclusions}.

\section{Computation and Monte Carlo implementation setup} \label{Sec:Setup}

In the context of the SMEFT, cross-sections can be decomposed in the following form:
\begin{align}
    d\sigma(\mu_R,\mu_F;\mu) &= d\sigma_{\rm SM}(\mu_R,\mu_F) \nonumber \\
    &+ \sum_i c_i(\mu) \, d\sigma_i(\mu_R,\mu_F;\mu) + \sum_{i \leq j} c_i(\mu) \, c_j(\mu) \, d\sigma_{ij}(\mu_R,\mu_F;\mu)\, + ...\,,
\label{xsecdecomp}
\end{align}
where the various Wilson coefficients are denoted by $c_i$, and the explicit dependence on nonphysical scales has been highlighted. In particular, $\mu_R$ denotes the SM renormalisation scale, $\mu_F$ the factorisation scale, and $\mu$ the EFT renormalisation scale. 

It is worth noting that the $\mu$ dependence of $d\sigma$ enters through the Wilson coefficients at all perturbative orders, and through the $d\sigma_{i\cdots}$ at one-loop and beyond. In particular, if SMEFT corrections are only considered at the tree level, the only $\mu$ dependence is through the RG flow of Wilson coefficients. \smallskip

The RGE of the SMEFT reads:
\begin{equation}
    \frac{dc_i(\mu)}{d\log\mu} = \bm{\gamma}_{ij} \,c_j(\mu),
\label{eq:rge}
\end{equation} 
with $\bm{\gamma}_{ij}$ the anomalous dimension matrix. 
We focus here on the QCD-induced part of the running, i.e. we ignore terms in the anomalous dimension matrix which are not proportional to $\alpha_s$. The $\bm{\gamma}$ matrix is then expanded in $\alpha_s$ as:
\begin{equation}
    \bm{\gamma}_{ij} = \sum_{k=1} \bigg(\frac{\alpha_s}{4\pi}\bigg)^k \bm{\gamma}_{ij}^{ \text{\tiny \rm QCD,k}}
\end{equation}

Due to the large value of $\alpha_s$, we expect $\bm{\gamma}_{ij}^{\text{\tiny \rm QCD,1}}$ to typically give the leading contribution to the running and mixing of the Wilson coefficients at present hadron collider energies\footnote{Exceptions are known, for instance $t\bar{t}W$ and 4-top~\cite{Frederix:2017wme,Aoude:2022deh} production receive sizable EW contributions.}.

The solution to the RGE equation \eqref{eq:rge} is given by:  
\begin{align}
    c_i(\mu) = \bm{\Gamma}_{ij}(\mu,\mu_0) \, c_j(\mu_0),
\end{align}
where $\mu_0$ is a reference scale. The $\bm{\Gamma}$ matrix can be evaluated order by order in $\alpha_s$, at order 1 it reads:
\begin{align}
    \bm{\Gamma}^{ \text{\tiny \rm QCD,1}}(\mu,\mu_0) \equiv \textrm{exp} \left( \int_{\mu_0}^\mu \frac{\alpha_s(\mu')}{4 \pi \mu'} \, d\mu' \  \bm{\gamma}^{ \text{\tiny \rm QCD,1}}\right). \label{eq:rge_solution}
\end{align}

The computation described above forms the basis of our Monte Carlo implementation, which takes:
\begin{align}
    \bm{\Gamma}^{ \text{\tiny \rm QCD,1}}(\mu,\mu_0) = \textrm{exp} \left( \frac{1}{2\beta_0} \log \frac{\alpha_s(\mu_0)}{\alpha_s(\mu)} \bm{\gamma}^{ \text{\tiny \rm QCD,1}} \right), \quad \beta_0 = 11 - \frac{2}{3} n_f, \label{eq:rge_montecarlo}
\end{align}
obtained by using the one-loop accurate expression for $\alpha_s(\mu)$ in \eqref{eq:rge_solution}; $n_f$ represents the number of light flavours. 

The running of $\alpha_s$ is itself modified by SMEFT operators \cite{Jenkins:2013zja}, however, given the present bounds, this effect is completely negligible for our purposes, and we will use the SM running of $\alpha_s$ in our MC implementation.

In practice the numerical values of Wilson coefficients are set by the user at a given scale $\mu_0$; an EFT renormalisation scale ($\mu$) can then be selected, and the code will automatically perform the corresponding running, using Eq.\eqref{eq:rge_montecarlo}, to the desired value of $\mu$. The EFT scale can be fixed, e.g. $\mu = m_{\rm top}$, or can be a chosen on a event-by-event basis, as a function of the event kinematics, e.g. $\mu = H_T/2$. While our discussion is based on the case of the SMEFT, the Monte Carlo implementation is generic and can be adapted to any scenario with running couplings. Further details on our implementation are given in App. \ref{sec:Gen_details}.

\section{RGE of top quark operators} \label{Sec:RGE_top}

To illustrate our implementation of RGE effects,  we consider the operators relevant for top quark pair production at a hadron collider. These operators will be grouped in two categories: 0/2-fermion operators and 4-fermion operators. We will focus on operators entering top pair production, and following the {\tt SMEFTatNLO} implementation \cite{Degrande:2020evl} we assume a $\text{U(2)}_q \otimes \text{U(2)}_u \otimes \text{U(3)}_d$ flavour symmetry, and CP conservation.\smallskip

In the following, we will present the anomalous dimension matrix $\bm{\gamma}^{ \text{\tiny \rm QCD,1}}$, entering the RGE at order $\alpha_s$, for the set of degrees of freedom discussed above. The RGE flow of the two subsets of operators, 0/2-fermion and 4-fermion, fully decouples, and the $\bm{\gamma}^{ \text{\tiny \rm QCD,1}}$ matrix is block diagonal.

We note that $\bm{\gamma}$ is a sparse matrix; the location of its zeros was observed in \cite{Alonso:2014rga} by looking at its holomorphic structure and later explained in \cite{Cheung:2015aba} from unitarity cuts in one-loop scattering amplitudes.
The zeros at two-loops were further explored in~\cite{Bern:2020ikv} using unitarity cuts and the formalism of~\cite{Caron-Huot:2016cwu}, which allows one to obtain the anomalous dimension at higher-loops from phase-space integrals of lower-loop form factors and amplitudes. This method have been continued to be explored in the SMEFT~\cite{Baratella:2020lzz,Baratella:2022nog}.
The zeros obtained in \cite{Alonso:2014rga,Cheung:2015aba} are assumed in our calculation. The non-zero entries, on the contrary, have been extracted independently of previous results, from the counterterms of the {\tt SMEFTatNLO} model \cite{Degrande:2020evl}, obtained within the {\tt NLOCT} framework \cite{Degrande:2014vpa}.

\subsection{Bosonic and two-quark operators}

Under our flavour assumption, there are two purely bosonic (0-fermion) operators to consider for top pair production:
\begin{equation}
    O_G = g_{S}f_{ABC} G^{A \, \nu}_\mu\, G^{B \, \rho}_\nu G^{C \, \mu}_\rho, \qquad O_{\varphi G} = \bigg( \varphi^\dagger \varphi - \frac{v^2}{2}\bigg) G_A^{\mu\nu} G^A_{\mu\nu}. \label{eq:zerof}
\end{equation}
The triple-gluon operator $O_G$ is already very well constrained by multijet observables \cite{Krauss:2016ely,Hirschi:2018etq}, and since $c_G$ is not induced by the RGE flow of any other Wilson coefficient, we will not consider it further.

In addition to \eqref{eq:zerof}, there are six 2-fermion operators \cite{Aguilar-Saavedra:2018ksv} but only the following four are running/mixing with other operators at order $\alpha_s$:
\begin{subequations}
\begin{align}
O_{t\varphi}  &= \bigg( \varphi^\dagger \varphi - \frac{v^2}{2}\bigg)  \bar{Q}\, t \,\tilde{\varphi} + \text{h.c.}, 
\qquad
O_{tG}  = i g_S(\bar{Q}\tau^{\mu\nu}T_A t)\tilde{\varphi} G^A_{\mu\nu} + \text{h.c.},\\
O_{tW} &= i(\bar{Q}\tau^{\mu\nu}\tau_I t)\tilde{\varphi} W^I_{\mu\nu} + \text{h.c.}, 
\qquad\,
O_{tB} =  i(\bar{Q}\tau^{\mu\nu} t)\tilde{\varphi} B_{\mu\nu} + \text{h.c.}, 
\end{align} 
\end{subequations}
where $\tau^{\mu \nu} = \frac{1}{2}[\gamma^{\mu}, \gamma^{\nu}]$. 
In the following, we will trade $c_{tB}$ for the linear combination:
\begin{equation}
    c_{tZ} \equiv -\sin\theta_W \, c_{tB}+ \cos\theta_W \, c_{tW},
\end{equation}
for physical convenience. \smallskip

We obtain the $\bm{\gamma}^{ \text{\tiny \rm QCD,1}}$ matrix in this sector:
\begin{align}
\label{eq:Gamma0and2F}
\bm{\gamma_{\text{0/2F}}^{\text{\tiny \rm QCD,1}}} = \frac{1}{3}
\begin{pmatrix}
 -24 & 96 y_t & 96 y_t^2 & 0 & 0\\
0 & -6 \beta_0 & 12y_t & 0 & 0  \\
0 & 0 & 4  & 0 & 0\\
0 & 0 & 8 g_2 & 8 & 0 \\
0 & 0 & 8 g_2 \cos \theta_W - \nicefrac{40}{3} \, g_1 \sin \theta_W & 0 & 8\\
\end{pmatrix}
\end{align}
where Wilson coefficients are ordered as $\bm{c} = \{c_{t\varphi} \, c_{\varphi \, G} \, c_{tG} \, c_{tW} \, c_{tZ} \}$. With $g_1$, $g_2$, $\theta_W$ we denote the EW SM parameters, with $y_t = \sqrt{2} \, m_{t, \text{pole}} / v$ the top Yukawa, and $\beta_0 = 11 - 2/3 \, n_f$.

As in \cite{Degrande:2020evl}, we defined the operators $O_G$ and $O_{tG}$ with appropriate powers of $g_S$. The extra factors of $g_S$ are consistently evolved according to their RGE flow. Redefinitions of the 0/2-fermion operators with factors of $y_t$, $g_1$, $g_2$, $g_S$, as done in \cite{Maltoni:2016yxb, Deutschmann:2017qum}, remove the corresponding parameters ($y_t$, $g$, $\beta_0$) from $\bm{\gamma_{\text{0/2F}}^{\text{\tiny \rm QCD,1}}}$. We follow the normalisation of \texttt{SMEFTatNLO} \cite{Degrande:2020evl}, but leave the parametric dependence on SM parameters explicit in \eqref{eq:Gamma0and2F}, to allow for an easy conversion towards different conventions.

The anomalous dimension matrix $\bm{\gamma_{\text{0/2F}}^{\text{\tiny \rm QCD,1}}}$ was extracted in \cite{Jenkins:2013zja,Jenkins:2013wua,Alonso:2013hga}; the running of $O_{t\varphi}, O_{tG}, O_{tW}, O_{tB}$ also appeared in \cite{Zhang:2014rja}, and the running of $O_{\varphi G}, O_{t \varphi}, O_{tG}$ in \cite{Maltoni:2016yxb, Deutschmann:2017qum}. Once different normalisation conventions are accounted for, our results are consistent with the existing literature. \smallskip

\begin{figure}[h!]
	\centering
	\begin{subfigure}{.5\textwidth}
      \centering
      \includegraphics[width=\linewidth]{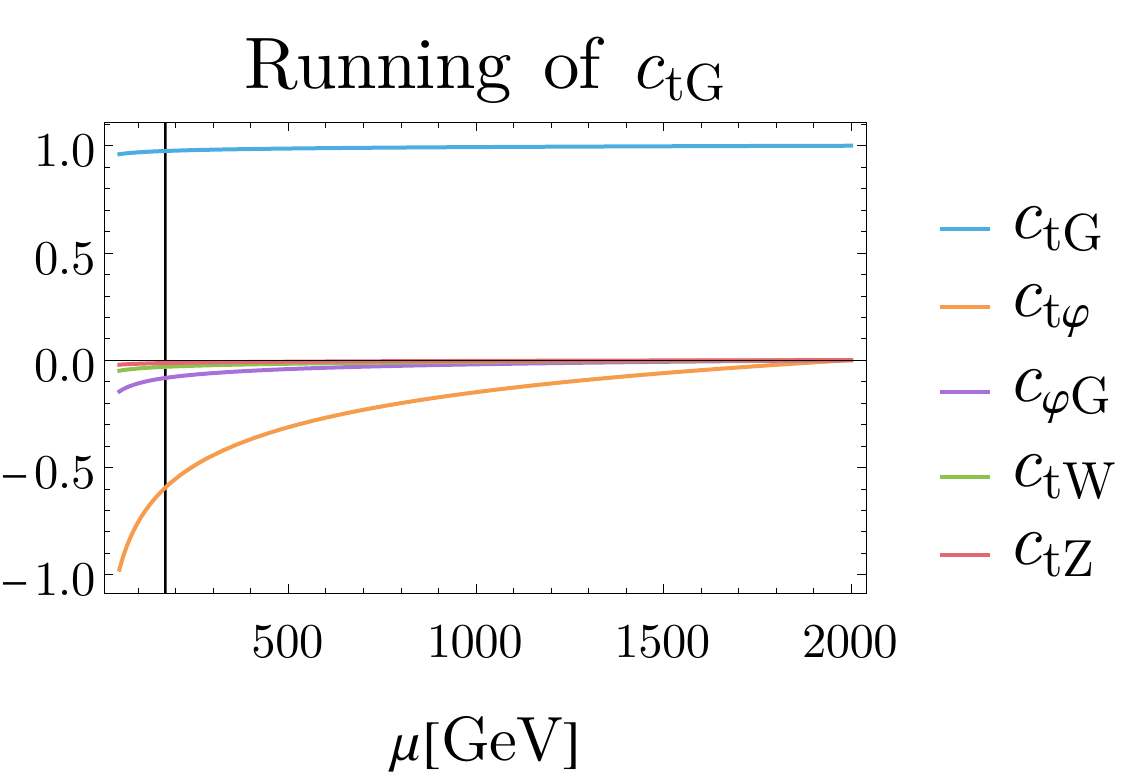}
    \end{subfigure}%
    \begin{subfigure}{.5\textwidth}
      \centering
      \includegraphics[width=\linewidth]{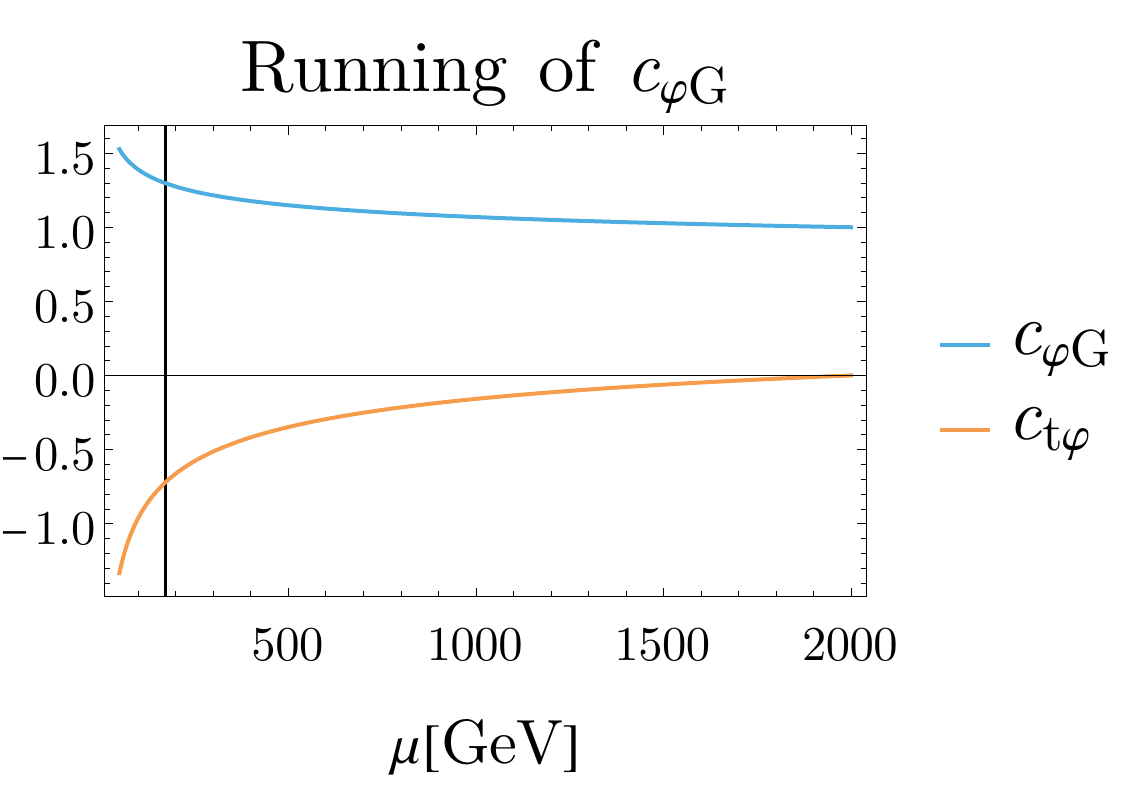}
    \end{subfigure}
	\caption{Renormalisation group running of the coefficients $c_{tG}$ (left) and $c_{\varphi G}$ (right), which are set to unity at 2 TeV and run as given by Eq.~\eqref{eq:Gamma0and2F}. The vertical grey line represents $m_{\rm top}$. The running of $c_{tG}$ mostly mixes into $c_{t \varphi}$, and to a lesser extent into all other coefficients in this sector. The running of $c_{\varphi G}$ induces $c_{t \varphi}$, and a change to its own value.}
	\label{fig:running_plots_0F}
\end{figure}

As an example, we show in Fig. \ref{fig:running_plots_0F} the RGE evolution of coefficients $c_{tG}$ and $c_{\varphi G}$. The coefficients are set to $1$ at $\mu_0 =$ 2 TeV and run down to lower scales. 
The choice of 2 TeV is somewhat arbirary, and represents a high scale where new physics resides. 

Each coefficient generated by the mixing is shown separately. We notice that the value of $c_{tG}$ gets reduced by a few percent by running down to $m_t$ and it mixes into several other operators. As obvious from the anomalous dimension matrix the largest coefficient induced by $c_{tG}$ is $c_{t\varphi}$. The running of $c_{\varphi G}$ is more pronounced (i.e. its value changes by about 30\%) and there is significant mixing into $c_{t\varphi}$.  

\subsection{Four-fermion operators}

In this Section, we present the RGE for the 4-fermion operators involving top quark fields. We organise 4-fermion operators in two categories, four-heavy (4H), involving four third generation quark fields, and two-light-two-heavy (2L2H), involving two third generation quark fields and two first or second generation quark fields. 

Under our flavour assumption, there are five 4H operators. In the LHC Top Working Group convention \cite{Aguilar-Saavedra:2018ksv,Degrande:2020evl}, these are given by $\lbrace O_{QQ}^{(1)} \, O_{Qt}^{(1)} \, O_{tt}^{(1)} \,  O_{QQ}^{(8)} \, O_{Qt}^{(8)} \rbrace$. 
The map between the degrees of freedom used here and those that appear in the Warsaw basis \cite{Grzadkowski:2010es} is given by:
\begin{subequations}
\begin{align}
c_{QQ}^{(1)} \equiv 2C_{qq}^{1(3333)} - \frac{2}{3}C_{qq}^{3(3333)}, \qquad c_{Qt}^{(1)} \equiv C_{qu}^{1(3333)}, \\
 c_{tt}^{(1)} \equiv C_{uu}^{1(3333)}, \quad c_{QQ}^{(8)} \equiv 8C_{qq}^{3(3333)}, \quad c_{Qt}^{(8)} \equiv C_{qu}^{8(3333)}.
\end{align}
\end{subequations}

There are fourteen 2L2H operators, that we take to be $\lbrace O_{Qq}^{(1,1)} \, O_{Qq}^{(1,3)} \, O_{tu}^{(1)} \, O_{td}^{(1)} \, O_{Qu}^{(1)} \, O_{Qd}^{(1)} \, O_{tq}^{(1)} \rbrace$ and $\lbrace O_{Qq}^{(8,1)} \, O_{Qq}^{(8,3)} \, O_{tu}^{(8)} \, O_{td}^{(8)} \, O_{Qu}^{(8)} \, O_{Qd}^{(8)} \, O_{tq}^{(8)} \rbrace$. The relations between the 2L2H degrees of freedom we use and those in the Warsaw basis are: 
\begin{subequations}
\begin{align}
&c_{Qq}^{(1,1)} \equiv C_{qq}^{1,(ii33)} + \frac{1}{6}C_{qq}^{1,(i33i)}+\frac{1}{2}C_{qq}^{3(i33i)} \ {\scriptstyle i=1,2}, \quad  c_{Qq}^{(8,1)} \equiv C_{qq}^{1(i33i)} + 3C_{qq}^{3(i33i)}  \ {\scriptstyle i=1,2}, \label{eq:conv1}\\
&c_{Qq}^{(1,3)} \equiv C_{qq}^{3,(ii33)} + \frac{1}{6} C_{qq}^{1,(i33i)} - \frac{1}{6} C_{qq}^{3(i33i)}  \ {\scriptstyle i=1,2}, \quad c_{Qq}^{(8,3)} \equiv C_{qq}^{1(i33i)} - C_{qq}^{3(i33i)}  \ {\scriptstyle i=1,2}, \\
&c_{tu}^{(1)} \equiv C_{uu}^{(ii33) } + \frac{1}{3}C_{uu}^{(i33i)}  \ {\scriptstyle i=1,2}, \hspace{31mm} c_{tu}^{(8)} \equiv 2C_{uu}^{(i33i)}  \ {\scriptstyle i=1,2}, \label{eq:conv1a}  \\
&c_{td}^{(1)} \equiv C_{ud}^{1(33ii)}  \ {\scriptstyle i=1,2,3}, \hspace{45mm} c_{td}^{(8)} \equiv C_{ud}^{8(33ii)}  \ {\scriptstyle i=1,2,3}, \\
&c_{Qu}^{(1)} \equiv C_{qu}^{1(33ii)}  \ {\scriptstyle i=1,2}, \hspace{48mm} c_{Qu}^{(8)} \equiv C_{qu}^{8(33ii)}  \ {\scriptstyle i=1,2}, \\
&c_{Qd}^{(1)} \equiv C_{qd}^{1(33ii)}  \ {\scriptstyle i=1,2,3}, \hspace{46mm} c_{Qd}^{(8)} \equiv C_{qd}^{8(33ii)}  \ {\scriptstyle i=1,2,3}, \\
&c_{tq}^{(1)} \equiv C_{qu}^{1(ii33)}  \ {\scriptstyle i=1,2}, \hspace{48mm} c_{tq}^{(8)} \equiv C_{qu}^{8(ii33)}  \ {\scriptstyle i=1,2}. \label{eq:conv2}
\end{align}
\end{subequations}

To clarify the meaning of Equations \eqref{eq:conv1}-\eqref{eq:conv2}, taking $c_{Qd}^{(1)}$ as an example, we set $C_{qd}^{1(3311)} = C_{qd}^{1(3322)} = C_{qd}^{1(3333)}$, and call this common value $c_{Qd}^{(1)}$.

 Table \ref{tab:FourFermionsChirality} summarises the operators we consider in this Section. Operators are split in 2L2H and 4H as described above, then by their colour structure, singlet or octet, and additionally by their chirality. There is no octet 4H RR operator, since the would-be operator $O_{tt}^{(8)}$ reduces to our $O_{tt}^{(1)}$ due to $SU(3)$ identities.  

\begin{table}[h]
	\renewcommand{\arraystretch}{1.25}
	\centering
	\begin{tabular}{|c|ccc|ccc|}\hline
	 & \multicolumn{3}{c|}{Two-light-two-heavy} &  \multicolumn{3}{c|}{Four-heavy} \\ \hline\hline
	  & LL & RR & LR/RL & LL & RR & LR/RL \\ \hline
	 Singlet & $O^{(1,1)}_{Qq}$ $O^{(1,3)}_{Qq}$ & $O_{tu}^{(1)}$ $O_{td}^{(1)}$ & $O_{Qu}^{(1)}$ $O_{Qd}^{(1)}$ $O_{tq}^{(1)}$ & $O^{(1)}_{QQ}$ &  $O_{tt}^{(1)}$ & $O_{Qt}^{(1)}$ \\
	  Octet & $O^{(8,1)}_{Qq}$ $O^{(8,3)}_{Qq}$ & $O_{tu}^{(8)}$ $O_{td}^{(8)}$ & $O_{Qu}^{(8)}$ $O_{Qd}^{(8)}$ $O_{tq}^{(8)}$ & $O^{(8)}_{QQ}$ & -  & $O_{Qt}^{(8)}$ \\ \hline
	\end{tabular}
	\caption{List of four fermion operators involving top quark fields classified according to their flavour content, colour and chirality structure.}
	\label{tab:FourFermionsChirality}
\end{table}

In this sector, we order Wilson coefficients as:
\begin{align}
\bm{c} = \{c_{tt}^{(1)} \, \big| \, c_{Qq}^{(1,1)},c_{Qq}^{(1,3)}, c_{tu}^{(1)}, c_{td}^{(1)},c_{Qu}^{(1)}, c_{Qd}^{(1)}, &c_{tq}^{(1)}, c_{QQ}^{(1)}, 
c_{Qt}^{(1)} \, \big| \,  \\ 
& c_{Qq}^{(8,1)}, c_{Qq}^{(8,3)},c_{tu}^{(8)}, c_{td}^{(8)}, c_{Qu}^{(8)}, c_{Qd}^{(8)}, 
c_{tq}^{(8)}, c_{QQ}^{(8)},  c_{Qt}^{(8)}\}, \notag
\end{align}
where a line separates the colour-singlet from the colour-octet sectors, with $c_{tt}^{(1)}$ being a special case. We obtain the following anomalous dimension matrix:

\begin{align}
\bm{\gamma_{4\text F}^{\text{\tiny \rm QCD,1}}} = \frac{1}{3}
\begin{small}
\left(
\begin{array}{c|ccccccccc|ccccccccc}
\renewcommand*{\arraystretch}{2.5}
\nicefrac{44}{3} & 0 & 0 & 0 & 0 & 0 & 0 & 0 & 0 & 0 & 0 & 0 & \nicefrac{4}{3} & 2 & 0 & 0 & \nicefrac{8}{3} & 0 & \nicefrac{4}{3} \\ \hline
0 & 0 & 0 & 0 & 0 & 0 & 0 & 0 & 0 & 0 & 8 & 0 & 0 & 0 & 0 & 0 & 0 & 0 & 0 \\
0 & 0 & 0 & 0 & 0 & 0 & 0 & 0 & 0 & 0 & 0 & 8 & 0 & 0 & 0 & 0 & 0 & 0 & 0 \\
0 & 0 & 0 & 0 & 0 & 0 & 0 & 0 & 0 & 0 & 0 & 0 & 8 & 0 & 0 & 0 & 0 & 0 & 0 \\
0 & 0 & 0 & 0 & 0 & 0 & 0 & 0 & 0 & 0 & 0 & 0 & 0 & 8 & 0 & 0 & 0 & 0 & 0 \\
0 & 0 & 0 & 0 & 0 & 0 & 0 & 0 & 0 & 0 & 0 & 0 & 0 & 0 & -8 & 0 & 0 & 0 & 0 \\
0 & 0 & 0 & 0 & 0 & 0 & 0 & 0 & 0 & 0 & 0 & 0 & 0 & 0 & 0 & -8 & 0 & 0 & 0 \\
0 & 0 & 0 & 0 & 0 & 0 & 0 & 0 & 0 & 0 & 0 & 0 & 0 & 0 & 0 & 0 & -8 & 0 & 0 \\
0 & 0 & 0 & 0 & 0 & 0 & 0 & 0 & 0 & 0 & 0 & 0 & 0 & 0 & 0 & 0 & 0 & 8 & 0 \\
0 & 0 & 0 & 0 & 0 & 0 & 0 & 0 & 0 & 0 & 0 & 0 & 0 & 0 & 0 & 0 & 0 & 0 & -8 \\
\hline 
0 & 36 & 0 & 0 & 0 & 0 & 0 & 0 & 4 & 0 & 0 & 0 & 0 & 0 & 4 & 6 & 2 & \nicefrac{10}{3} & 2 \\
0 & 0 & 36 & 0 & 0 & 0 & 0 & 0 & 0 & 0 & 0 & -12 & 0 & 0 & 0 & 0 & 0 & 0 & 0 \\
8 & 0 & 0 & 36 & 0 & 0 & 0 & 0 & 0 & 0 & 0 & 0 & -6 & 6 & 4 & 0 & 8 & 0 & 4 \\
8 & 0 & 0 & 0 & 36 & 0 & 0 & 0 & 0 & 0 & 0 & 0 & 4 & -4 & 0 & 4 & 8 & 0 & 4 \\
0 & 0 & 0 & 0 & 0 & -36 & 0 & 0 & 4 & 0 & 8 & 0 & 2 & 0 & -34 & 6 & 0 & \nicefrac{10}{3} & 2 \\
0 & 0 & 0 & 0 & 0 & 0 & -36 & 0 & 4 & 0 & 8 & 0 & 0 & 2 & 4 & -32 & 0 & \nicefrac{10}{3} & 2 \\
8 & 0 & 0 & 0 & 0 & 0 & 0 & -36 & 0 & 0 & 4 & 0 & 4 & 6 & 0 & 0 & -32 & 0 & 4 \\
0 & 0 & 0 & 0 & 0 & 0 & 0 & 0 & 44 & 0 & 16 & 0 & 0 & 0 & 8 & 12 & 0 & -\nicefrac{16}{3} & 4 \\
8 & 0 & 0 & 0 & 0 & 0 & 0 & 0 & 4 & -36 & 8 & 0 & 4 & 6 & 4 & 6 & 8 & \nicefrac{10}{3} & -36 \\
\end{array}
\right)
\end{small}
\notag
\end{align}

\medskip

Excluding $c_{tt}^{(1)}$, that has no colour-octet partner, we note that the upper left block has only zero entries, meaning that singlets do not run under QCD. Further, the top-right block is diagonal, meaning singlets are only modified by their colour octet counterparts. This is known, and understood, in terms of unitarity cuts~\cite{Bern:2020ikv}. We also note the bottom-left block is almost diagonal, signalling that octets mix predominantly, but not exclusively, into their singlet partner. Finally, the bottom-right block is the busiest part of the matrix with several non-diagonal entries as colour octets both run and also mix with each other.

As mentioned above, the entries of $\bm{\gamma_{4\text F}^{\text{\tiny \rm QCD,1}}}$ have been obtained automatically, from the counterterms produced by {\tt NLOCT} \cite{Degrande:2014vpa} for the {\tt SMEFTatNLO} model \cite{Degrande:2020evl}.
The validity of such counterterms has been confirmed with the {\tt COLLIER} library \cite{Denner:2014gla}, from the cancellation of UV poles in a large variety of virtual amplitudes. The results of \cite{Jenkins:2013zja,Jenkins:2013wua,Alonso:2013hga} agree with ours, upon accounting for the different flavor symmetry, and for a different convention followed for operators with repeated currents \footnote{We thank Aneesh Manohar for clarifying the convention used in \cite{Jenkins:2013zja,Jenkins:2013wua,Alonso:2013hga}. This helped us resolve what originally seemed to be a disagreement between our findings and those of \cite{Alonso:2013hga}.}, that we describe in detail in Appendix \ref{sec:comparison}. \smallskip

\begin{figure}[!htb]
	\centering
	\begin{subfigure}{.5\textwidth}
      \centering
      \includegraphics[width=\linewidth]{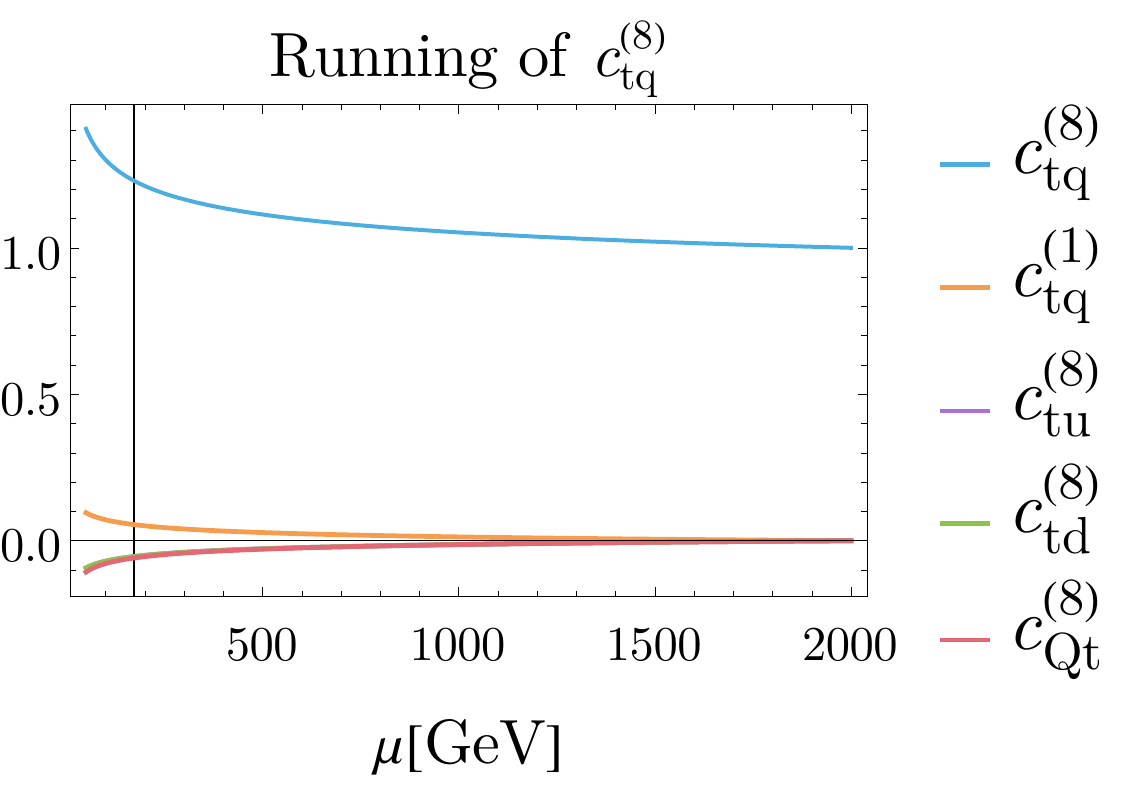}
    \end{subfigure}%
    \begin{subfigure}{.5\textwidth}
      \centering
      \includegraphics[width=\linewidth]{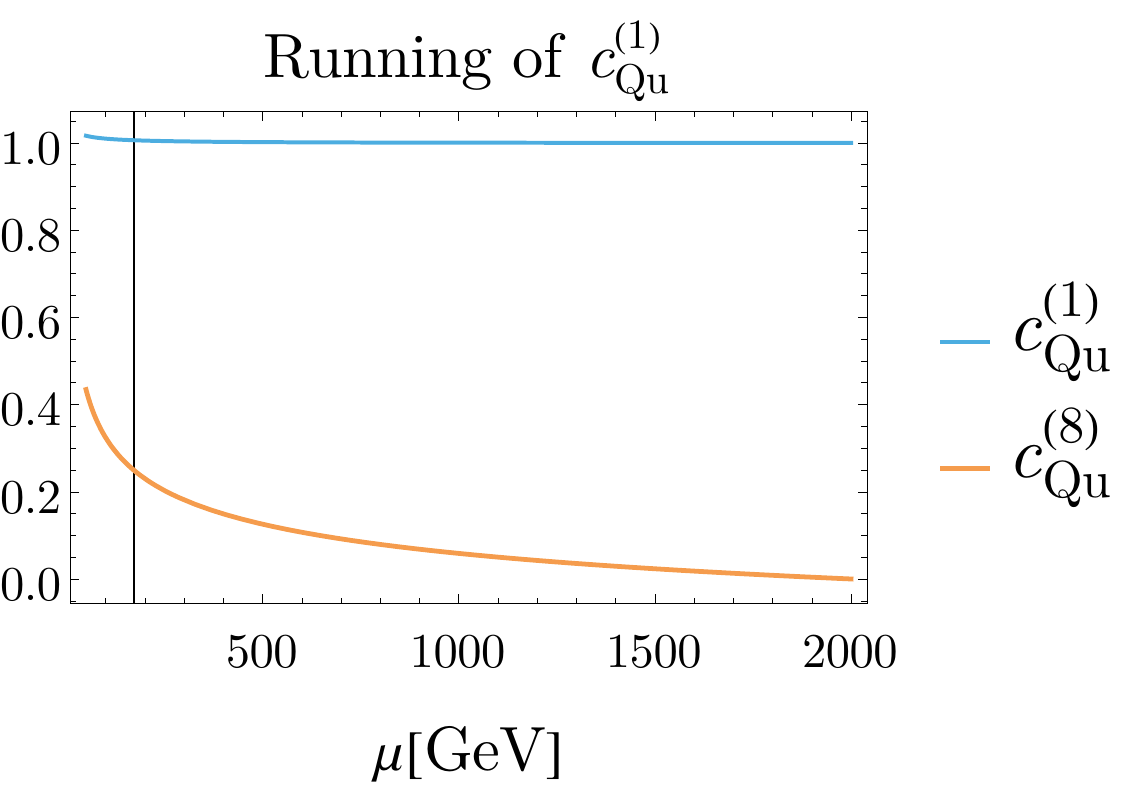}
    \end{subfigure}
	\caption{Renormalisation group running of the coefficients $c_{tQ}^{(8)}$ (left) and $c_{Qu}^{(1)}$ (right), which are set to unity at 2 TeV. The vertical grey line shows $m_{\rm top}$. Curves that do not deviate appreciably from zero in the $\mu$ range we considered are not plotted. 
	}
	\label{fig:running_plots_4F}
\end{figure}

In order to illustrate the impact of running and mixing, in Fig.~\ref{fig:running_plots_4F} we show as an example the running of $c_{tq}^{(8)}$ and $c_{Qu}^{(1)}$ with initial condition $c(2 \, \text{TeV}) = 1$. Each new operator coefficient induced by the mixing is shown separately in the plots. The two coefficients we have chosen to plot are representative of the general behaviour, singlets operators mostly run into their colour partner, while octets produce an intricate mixing pattern. The size of RGE shifts to Wilson coefficients when running from $2 \, \text{TeV}$ to $m_{\rm top}$  ranges from a few percent to around 20\% depending on the operator.

Whilst the anomalous dimension matrix gives a clear indication of the operator running and mixing, to explore the physical impact of RGE effects we will consider the top pair production cross-section in the next Section.

\section{Results for top pair production at the LHC} \label{Sec:Results_LHC}

The numerical effects of RGE on top-pair production can be explored in \texttt{Madgraph5\_aMC@NLO} \cite{Alwall:2014hca}. Our implementation is public, available from version 3.4.0, and further details are given in App. \ref{sec:Gen_details}.  In this Section, we study cross-sections and differential distributions for each operator and different scale choices, as explained in the following. 

\subsection{Scale choices}

The effects of running and mixing can be quantified by using different settings for the SMEFT renormalisation scale $\mu$. As discussed in Section \ref{Sec:Setup}, scale choices can be either considered fixed or dynamical. In the former case, $\mu$ is constant, and the SM renormalization and the factorisation scales can also be fixed, or allowed to be dynamical. In the latter case, $\mu$ is a dynamically-evaluated function of the final state's momenta computed for each phase space point. 

\subsection{Cross-section results}

We first explore the linear contributions to the cross-section, which correspond to the second term of Eq. \eqref{xsecdecomp},  $\sum_i c_i \, d\sigma_i$, induced by a set of operator coefficients defined at a scale $\mu_0$ and run to a different scale $\mu$. In Table \ref{tab:Linear_xsec} we show results for the SM-SMEFT linear interference for $p \, p \to t \, \bar t$ at $\sqrt{s} = 13 \, \text{TeV}$ using the NNPDF2.3 LO parton distribution function with $\alpha_s(m_Z) = 0.130$ \cite{Ball:2012cx}, for 0/2-fermion and 4-fermion operators respectively, evaluated at the four choices of scales: $(\mu_0 = 2 \, \text{TeV}, \mu = m_{\rm top})$, $(\mu_0 = 2 \, \text{TeV}, \mu = H_T/2)$, $(\mu_0 = m_{\rm top}, \mu = m_{\rm top})$, and $(\mu_0 = m_{\rm top}, \mu = H_T/2)$, at Leading Order in QCD and EW.  In addition to the EFT scales $\mu_0$ and $\mu$, Table \ref{tab:Linear_xsec} considers two possible choices of SM renormalisation and factorisation scales $\mu_R$ and $\mu_F$, the constant $m_{\rm top}$, and the dynamical $H_T/2$. The comparison between the first and last column of Table \ref{tab:Linear_xsec} gives an assessment of the spread between the total absence of RGE flow, and the now state-of-the-art in terms of including RGE effects.

\begin{table}[h]
	\centering
	\scalebox{0.8}
	{
	\begin{tabular}{|c|rr|rr|rr|r|} \hline \hline
	        \multirow{4}{*}{Operator} & \multicolumn{6}{c|}{Cross-section [pb]} & \\ \cline{2-7}
			& \multicolumn{2}{c|}{$\mu_R=\mu_F=m_{\rm top}$} 
			& \multicolumn{2}{c|}{$\mu_R=\mu_F=H_T/2$}&  \multicolumn{2}{c|}{$\mu_R=\mu_F=H_T/2$}&
			\multirow{3}{*}{$\frac{(\text{fix.}-\text{dyn.})}{\text{dyn.}}[\%]$}\\
			& $\mu_0=m_{\rm top}$  & $\mu_0$ = 2 TeV & $\mu_0=m_{\rm top}$  & $\mu_0$ = 2 TeV & $\mu_0=m_{\rm top}$   & $\mu_0$ = 2 TeV &\\
			& $\mu = m_{\rm top}$  & $\mu= m_{\rm top}$  & $\mu = m_{\rm top}$  & $\mu= m_{\rm top}$  & $\mu=H_T/2$  & $\mu=H_T/2$ &\\ \hline
$c_{t \varphi}$ & $0.000\ $ & $0.000\ $ & $0.000\ $ & $0.000\ $ & $0.000\ $ & $0.000\ $ & $ - \ \ $ \\
$c_{\varphi G}$ & $-4.343\ $ & $-5.781\ $ & $-3.992\ $ & $-5.302\ $ & $-3.842\ $ & $-5.112\ $ & $3.7 \ \% \ \ $ \\
$c_{tG}$ & $45.268\ $ & $44.431\ $ & $41.395\ $ & $40.649\ $ & $41.443\ $ & $40.727\ $ & $-0.2 \ \% \ \ $ \\
$c_{tW}$ & $0.128\ $ & $0.121\ $ & $0.125\ $ & $0.119\ $ & $0.126\ $ & $0.119\ $ & $-0.8 \ \% \ \ $ \\
$c_{tZ}$ & $-0.068\ $ & $-0.065\ $ & $-0.068\ $ & $-0.064\ $ & $-0.068\ $ & $-0.065\ $ & $-0.9 \ \% \ \ $ \\
\hline
$c_{Qq}^{(8,3)}$ & $0.350\ $ & $0.355\ $ & $0.324\ $ & $0.329\ $ & $0.322\ $ & $0.328\ $ & $0.4 \ \% \ \ $ \\
$c_{Qq}^{(8,1)}$ & $1.732\ $ & $1.594\ $ & $1.601\ $ & $1.475\ $ & $1.626\ $ & $1.496\ $ & $-1.4 \ \% \ \ $ \\
$c_{Qu}^{(8)}$ & $1.039\ $ & $1.222\ $ & $0.962\ $ & $1.133\ $ & $0.938\ $ & $1.095\ $ & $3.5 \ \% \ \ $ \\
$c_{tq}^{(8)}$ & $1.730\ $ & $2.047\ $ & $1.601\ $ & $1.896\ $ & $1.560\ $ & $1.833\ $ & $3.4 \ \% \ \ $ \\
$c_{Qd}^{(8)}$ & $0.706\ $ & $0.744\ $ & $0.656\ $ & $0.690\ $ & $0.654\ $ & $0.682\ $ & $1.1 \ \% \ \ $ \\
$c_{tu}^{(8)}$ & $1.041\ $ & $0.999\ $ & $0.962\ $ & $0.923\ $ & $0.969\ $ & $0.929\ $ & $-0.6 \ \% \ \ $ \\
$c_{td}^{(8)}$ & $0.708\ $ & $0.610\ $ & $0.657\ $ & $0.567\ $ & $0.673\ $ & $0.582\ $ & $-2.7 \ \% \ \ $ \\
\hline
$c_{Qq}^{(1,3)}$ & $0.434\ $ & $0.348\ $ & $0.423\ $ & $0.344\ $ & $0.439\ $ & $0.359\ $ & $-4.3 \ \% \ \ $ \\
$c_{Qq}^{(1,1)}$ & $0.119\ $ & $-0.286\ $ & $0.117\ $ & $-0.258\ $ & $0.189\ $ & $-0.191\ $ & $35.0 \ \% \ \ $ \\
$c_{Qu}^{(1)}$ & $0.075\ $ & $0.350\ $ & $0.074\ $ & $0.328\ $ & $0.032\ $ & $0.279\ $ & $17.6 \ \% \ \ $ \\
$c_{tq}^{(1)}$ & $0.087\ $ & $0.545\ $ & $0.085\ $ & $0.509\ $ & $0.016\ $ & $0.428\ $ & $19.0 \ \% \ \ $ \\
$c_{Qd}^{(1)}$ & $-0.026\ $ & $0.150\ $ & $-0.026\ $ & $0.138\ $ & $-0.054\ $ & $0.108\ $ & $27.1 \ \% \ \ $ \\
$c_{tu}^{(1)}$ & $0.149\ $ & $-0.100\ $ & $0.145\ $ & $-0.085\ $ & $0.189\ $ & $-0.043\ $ & $97.1 \ \% \ \ $ \\
$c_{td}^{(1)}$ & $-0.050\ $ & $-0.211\ $ & $-0.049\ $ & $-0.198\ $ & $-0.020\ $ & $-0.173\ $ & $14.6 \ \% \ \ $ \\
\hline
$c_{QQ}^{(8)}$ & $-0.019\ $ & $-0.095\ $ & $-0.021\ $ & $-0.091\ $ & $-0.008\ $ & $-0.078\ $ & $16.1 \ \% \ \ $ \\
$c_{Qt}^{(8)}$ & $-0.005\ $ & $-0.169\ $ & $-0.006\ $ & $-0.158\ $ & $0.017\ $ & $-0.128\ $ & $23.2 \ \% \ \ $ \\
\hline
$c_{QQ}^{(1)}$ & $-0.031\ $ & $-0.107\ $ & $-0.033\ $ & $-0.102\ $ & $-0.017\ $ & $-0.090\ $ & $13.0 \ \% \ \ $ \\
$c_{Qt}^{(1)}$ & $-0.018\ $ & $-0.038\ $ & $-0.019\ $ & $-0.038\ $ & $-0.020\ $ & $-0.033\ $ & $17.3 \ \% \ \ $ \\
\hline
$c_{tt}^{(1)}$ & $0.000\ $ & $-0.179\ $ & $0.000\ $ & $-0.166\ $ & $0.032\ $ & $-0.137\ $ & $21.3 \ \% \ \ $ \\
\hline \hline
	\end{tabular}
	}
	
	\caption{ Cross-sections for $p \, p \to t \, \bar t$ at $\sqrt{s} = 13 \, \text{TeV}$ at linear order in $c/\Lambda^2$ and at LO QCD and LO EW for various choices of SM renormalisation scale $\mu_R$, factorisation scale $\mu_F$, and EFT scale $\mu$. The associated Wilson coefficients is set to $1$ at $\mu_0$ and run to $\mu$ using the RGE we extracted; $\Lambda$ is set to 2 TeV. The Monte Carlo uncertainty is beyond the quoted digits. From top to bottom: 0/2-fermions, 4-fermions colour-octet 2L2H,  4-fermions colour-singlet 2L2H,  4-fermions colour-octet 4H, 4-fermions colour-singlet 4H, $c_{tt}^{(1)}$. The last column refers to the difference between $\mu = m_{\text{top}}$ and $\mu = H_T/2$ for $\mu_0 = 2 \, \text{TeV}$ and $\mu_R=\mu_F=H_T/2$.}
	\label{tab:Linear_xsec}
\end{table}

As it is evident, the spread between these two extremes is often producing a correction of order one to the interference cross-sections,
signaling the importance of including running and mixing effects. 
The spread between $\mu=m_{\text{top}}$ and $\mu=H_T/2$ can be considered as a first estimate of theoretical uncertainty. To assess the shift induced by different choices of running in the SMEFT sector, we compare the $\mu = m_{\rm top}$ and $\mu = H_T/2$ cross sections, keeping $\mu_0 = 2 \, \text{TeV}$ and $\mu_R = \mu_F = H_T/2$, i.e. the fifth and seventh columns of the table. Apart from $c_{t \varphi}$, that only enters top production at one loop, we note that the spread is small for the rest of 0/2-fermion operators, amounting to a few percent difference at most. The same holds true for the two-light-two-heavy colour-octets. The 2L2H colour-singlet cross-sections, on the other hand, show a dramatic change. This effect is due to the small initial SM-SMEFT interference (only with the EW SM amplitude), that is enhanced significantly by the respective colour-octets induced by the RGE, for which the interference is much larger. Typical shifts between using a fixed and a dynamical scale are of order $20 \%$, with outliers, such as $c_{Qq}^{(1,1)}$ and $c_{tu}^{(1)}$, deviating much more, up to $100 \%$. A similar effect happens in the case of 4H operators, for both colour-octet and singlets, that can only enter $b \bar b \to t \bar t$ (at LO), and thanks to renormalisation group mixing induce 2L2H operators, that escape the PDF suppression. This typically produces a spread of order $20 \%$ between different EFT running choices. This comparison not only highlights the importance of considering RGE effects at all, but shows how the choice of SMEFT renormalisation scale should be carefully considered, just like it is commonly done for the SM.   \smallskip

\begin{figure}[!htb]
	\centering
	\begin{subfigure}{.50\textwidth}
      \centering
      \includegraphics[width=\textwidth]{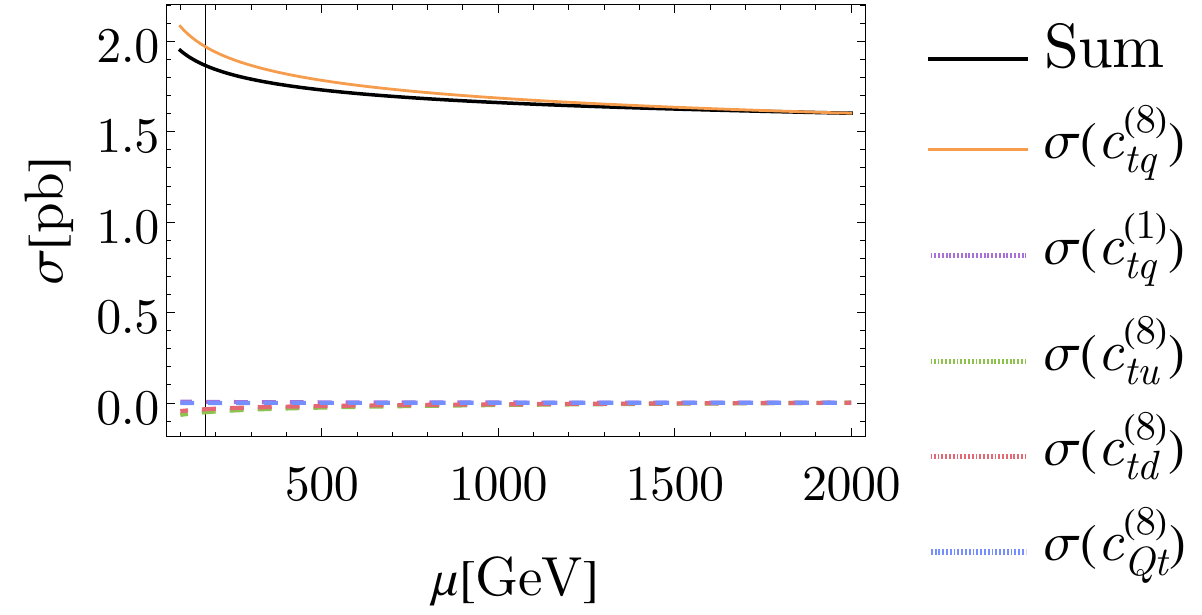}
    \end{subfigure}%
    \begin{subfigure}{.50\textwidth}
      \centering
      \includegraphics[width=\textwidth]{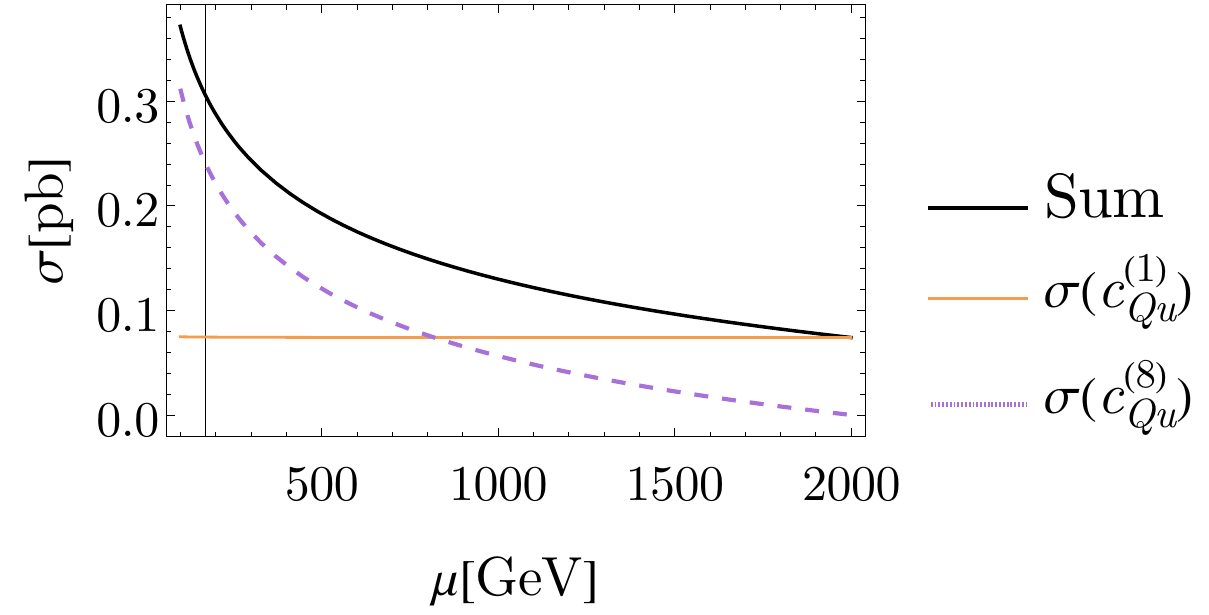}
    \end{subfigure}
	\caption{Renormalisation group flow of the tree-level (QCD+EW) $t\bar{t}$ interference cross-section induced by $c_{tq}^{(8)}(\mu_0=2 \textrm{TeV})=1$ (left) and $c_{Qu}^{(1)}(\mu_0=2 \textrm{TeV})=1$ (right), which are set to unity at 2 TeV. $\Lambda$ is set to 2 TeV. The contributions proportional to each Wilson coefficient are drawn separately in colour, and the total cross-section is drawn in black. The vertical grey line shows $m_{\rm top}$.	}
	\label{fig:xsec_running}
\end{figure}

The results shown in Tab.~\ref{tab:Linear_xsec} can be further visualised by decomposing, operator by operator, the total cross-section into the terms induced by the original operator, and by the mixing into others. As an example, in Fig.~\ref{fig:xsec_running} we show the SM-SMEFT interference cross-section for $p \, p \to t \, \bar t$ at $\sqrt{s} = 13 \, \text{TeV}$ split in individual contributions from various operators, obtained from setting $c_{tq}^{(8)}$ and $c_{Qu}^{(1)}$ to 1 at $\mu_0 = 2 \, \text{TeV}$ and running down to lower scales.

\smallskip

\subsection{Differential Distributions}
\begin{figure}[H]
	\centering
	\includegraphics[width=.6\textwidth]{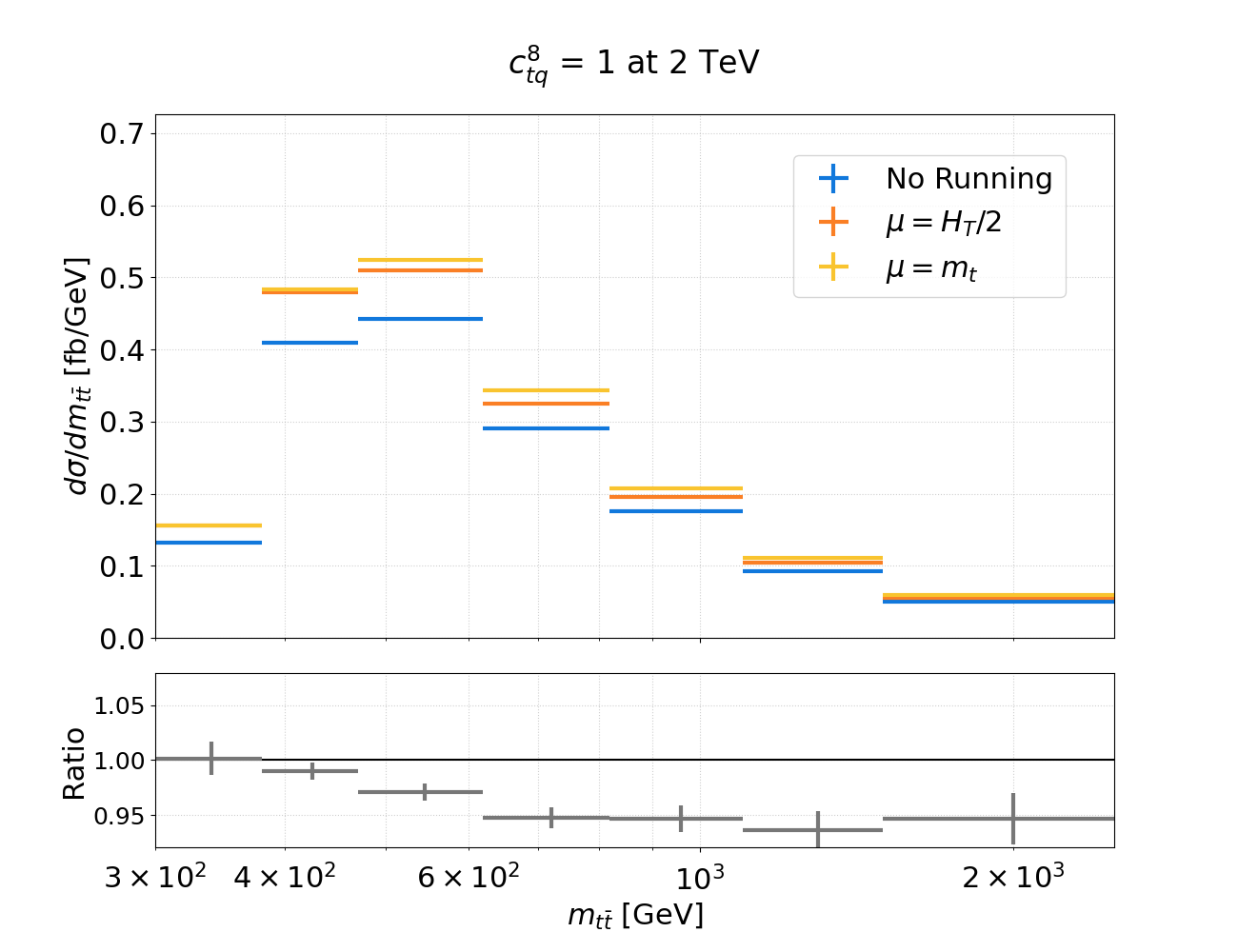}
	\includegraphics[width=.6\textwidth]{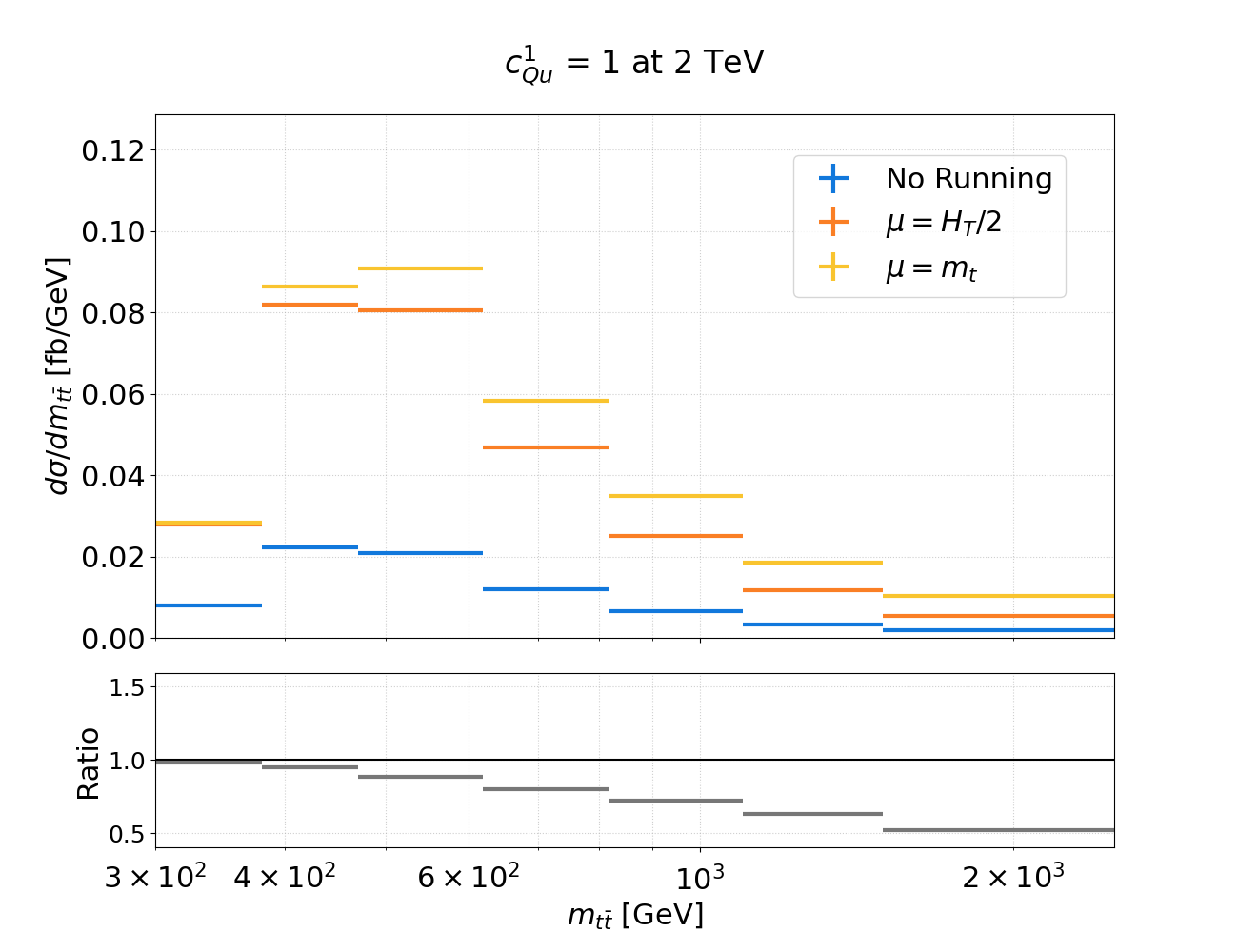}
	\caption{Linear interference contribution at LO QCD and LO EW to the $t \bar t$ invariant mass differential cross-section for $pp \to t \bar t$ at $\sqrt{s}=13 \, \text{TeV}$ induced by the 2L2H colour-octet operator $\mathcal O_{tq}^{(8)}$ (top) and by the 2L2H colour-singlet operator $\mathcal O_{Qu}^{(1)}$ (bottom), under the scale choices $\mu=m_{\rm top}$ and $\mu = H_T/2$; $\Lambda$ is set to 2 TeV. The coefficients are set to $1$ at $\mu_0 = 2 \, \text{TeV}$ and RGE evolved. Results obtained without running are shown for comparison. The SM renormalisation and factorisation scales $\mu_R$ and $\mu_F$ are set to $H_T/2$. The bottom plot shows the ratio between $\mu = H_T/2$ and $\mu=m_{\rm top}$; uncertainty is Monte Carlo.}
	\label{fig:mtt_scales}
\end{figure}

In addition to the total $t \bar t$ cross-section, we study the impact of operator running and mixing on the distribution of the top pair invariant mass $m_{t \bar t}$.
In the following plots, we show the distributions for fixed and dynamical scales. In the fixed scale case, as above, the relevant coefficient is set to one at $2 \ \text{TeV}$ and RGE evolved down to the top mass; in the dynamical scale case, the initial condition is the same, but the RGE evolution stops at the phase-space dependent point $\mu = H_T/2$. We show results for two 4-fermion operators, one colour-octet 2L2H and one colour-singlet 2L2H, in Fig. \ref{fig:mtt_scales}. 

The impact of a different scale choice is moderate for the colour-octet 2L2H operator reaching at most 10\%, as already discussed above, while it amounts to a significant, $\mathcal O(50\%)$, shift for the 2L2H colour-singlet operator. As expected, the difference between our two scale choices is larger for the higher energy bins, where $H_T/2 \gg m_{t \bar t}$, while the two scales coincide at threshold as shown in the inset of Fig. \ref{fig:mtt_scales}. We provide additional plots, similar to Figure \ref{fig:mtt_scales}, for other 2L2H and 4H operators in Appendix \ref{sec:Additional_differential}.

\subsection{Comparison of NLO with the RGE-evolved LO}
\label{sec:RGEvsNLO}

\begin{figure}[H]
	\centering
	
	 \begin{subfigure}{.5\textwidth}
      \centering
      \includegraphics[width=\linewidth]{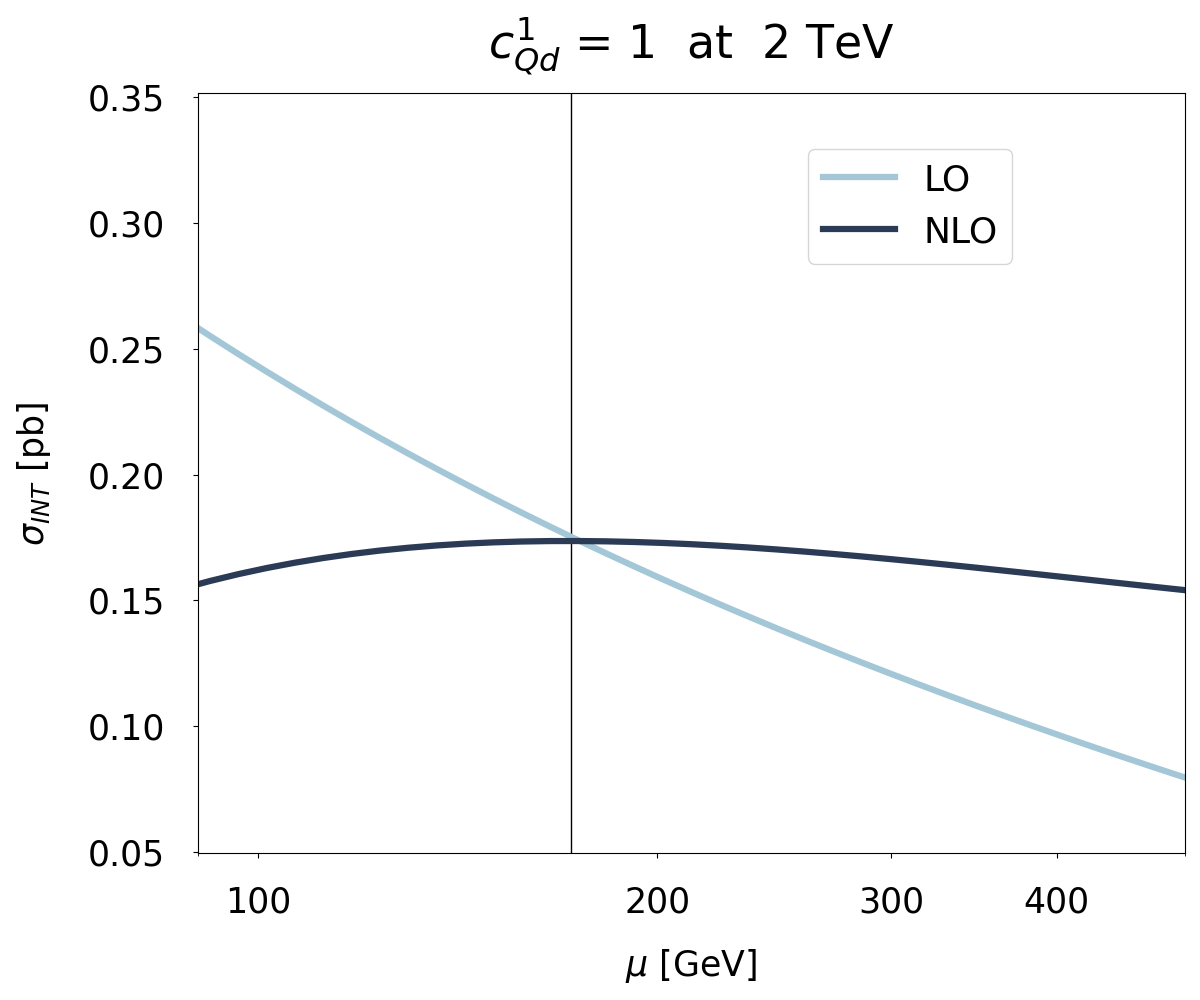}
    \end{subfigure}%
    \begin{subfigure}{.5\textwidth}
      \centering
      \includegraphics[width=\linewidth]{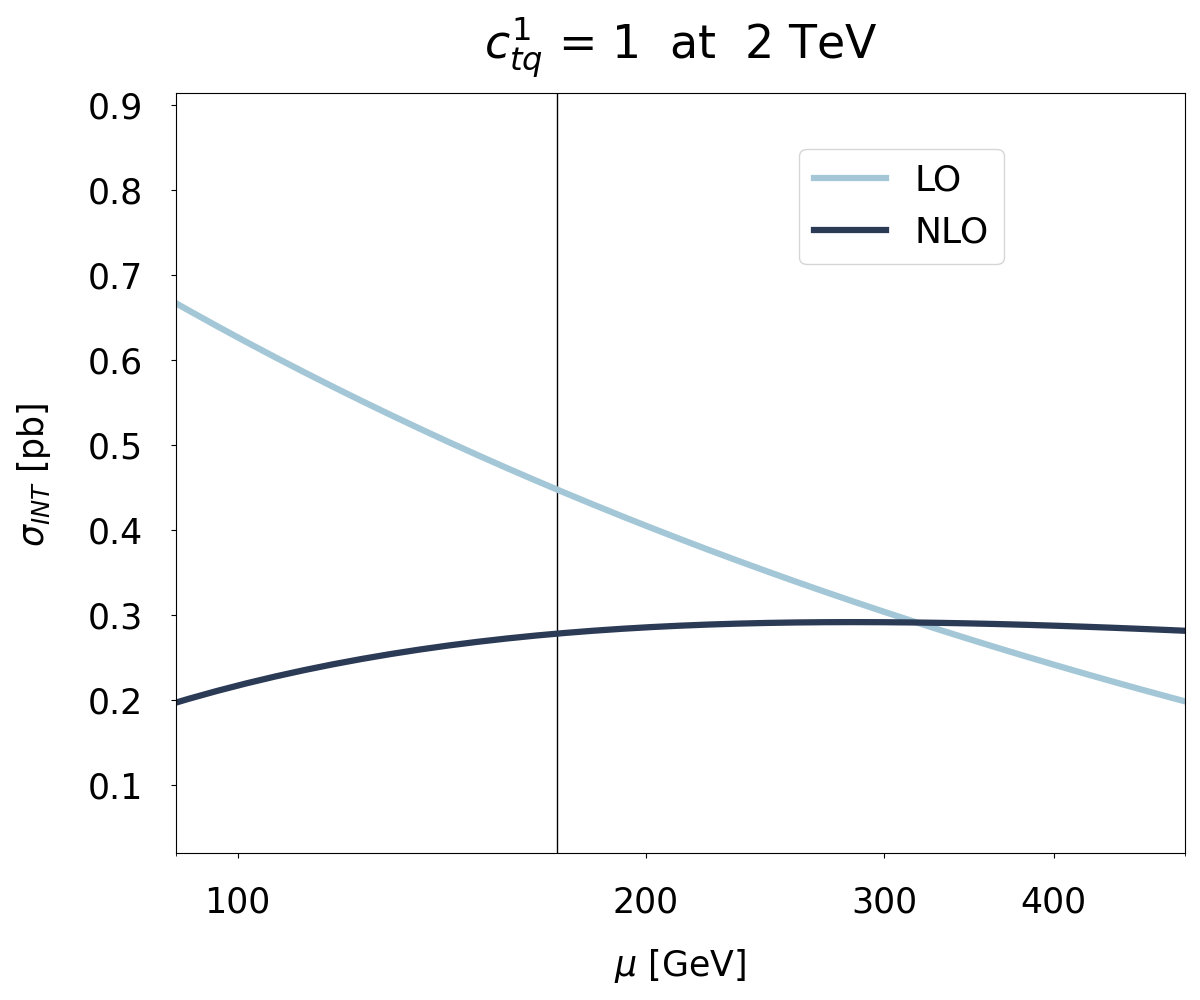}
    \end{subfigure}
	\begin{subfigure}{.5\textwidth}
      \centering
      \includegraphics[width=\linewidth]{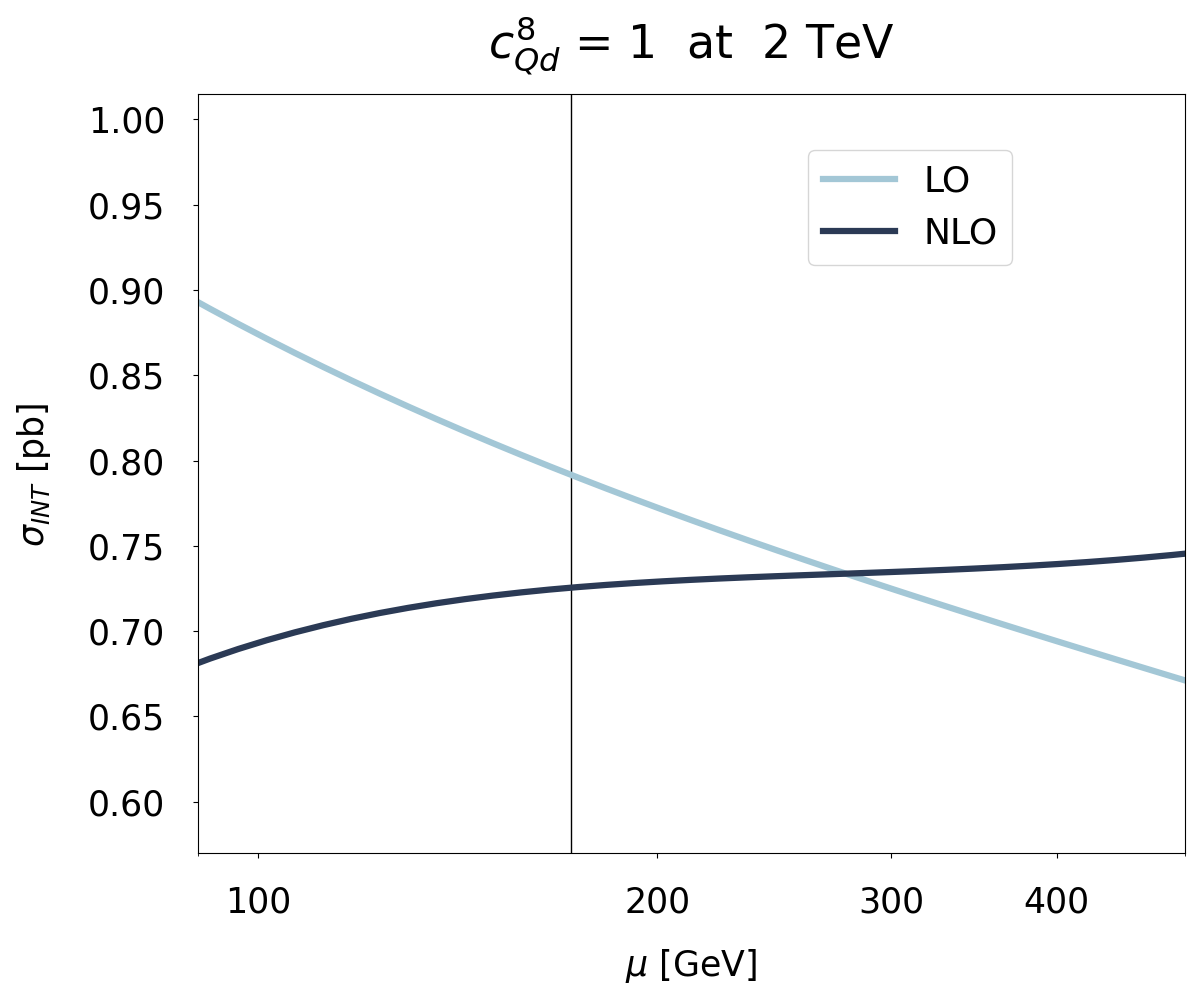}
    \end{subfigure}%
    \begin{subfigure}{.5\textwidth}
      \centering
      \includegraphics[width=\linewidth]{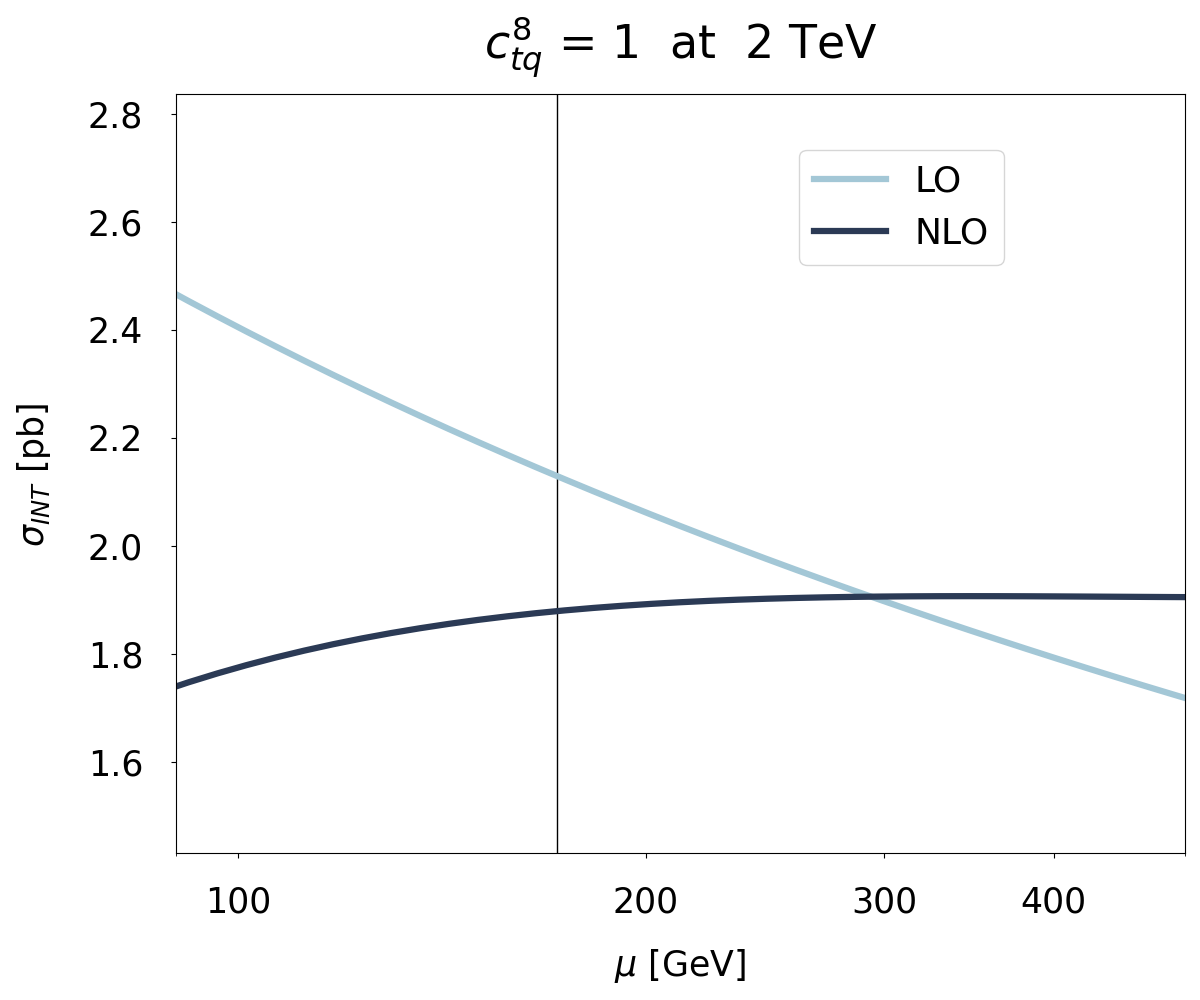}
    \end{subfigure}

	\caption{LO vs. NLO comparison for the SM-SMEFT interference cross-section for top pair production for four selected two-light-two-heavy operators. The Wilson coefficients start at $c(2 \ \text{TeV}) = 1$ and are RGE-evolved down to a lower scale $\mu$. $\Lambda$ is set to 2 TeV. The SM renormalisation scale $\mu_R$ is also set equal to $\mu$, while the factorisation scale is fixed to $\mu_F = m_{\text{top}}$. A vertical line is drawn at $m_\text{top}$.
	}
	\label{fig:NLOvsLOcomparison}
\end{figure}

The interference cross-sections 
obtained in previous Sections are Leading-Log (LL) improved LO results, and thus do not contain all the information that would be present in a full NLO calculation. In this Section, we aim to determine if LO+LL results can serve as a proxy for results at NLO. 
To do so, we evaluate the LO and NLO QCD cross-section for top pair production at $\sqrt{s} = 13 \ \text{TeV}$ with, as above, the Wilson coefficients defined as unity at $\mu_0 = 2 \ \text{TeV}$ and run to a lower scale $\mu$. We show the comparison for four selected operators in Fig. \ref{fig:NLOvsLOcomparison}.  Both the LO and NLO interferences are evolved under the one-loop RGE we presented in the previous sections.

We note that for a formal NLO accuracy the 2-loop anomalous dimension matrix would be needed, but this result is not yet available, therefore we are only able to employ the one-loop anomalous dimension for our comparisons. 

Two features are evident from Fig. \ref{fig:NLOvsLOcomparison}: first, the NLO cross-section is significantly more stable with respect to scale variations than the LO one, as expected. Second, the scale variation of LO seems to serve as a good proxy for uncertainty due to missing higher orders. Third, relevant for our comparison, RGE corrections usually improve the leading-order predictions, especially in the case of colour-singlet operators. Starting from the initial condition at the high scale $\mu_0$, the LO cross-section tends to run in the direction of its NLO value, growing if the NLO is larger and decreasing if the NLO is smaller. For instance, in the case of $\mathcal O_{Qd}^1$, the RGE-corrected LO cross-section and the NLO cross-section almost exactly agree at $\mu = m_\text{top}$.  
In most cases, an RGE evolution to some intermediate value $\mu \gtrsim m_\text{top}$ gives a relatively good agreement between the LO and NLO cross-sections.

It must be noted, however, that whilst RGE corrections to LO cross-sections improve their accuracy, they still do not capture the features of a full NLO calculation as they are missing the finite pieces present in the full one-loop computation which can potentially be large. Similar conclusions have been drawn previously in studies of Higgs \cite{Deutschmann:2017qum} and top-Higgs associated production \cite{Maltoni:2016yxb}.

We conclude this section by stressing that the computation of the two-loop anomalous dimension is desirable to promote the formal precision of SMEFT predictions to NLO. We note though that our computation and implementation is a first necessary step towards properly accounting for RGE effects in SMEFT predictions. Our implementation can be extended to include the two-loop anomalous dimension once this becomes available.

\section{Impact of RGE effects on constraints on the EFT}

\label{Sec:ToyFit}
In order to assess the impact of the scale choice and corresponding running and mixing of the operators on bounding the Wilson coefficients, we perform a toy fit for top pair production observables in the four-fermion sector of the SMEFT. Here we are not trying to set realistic bounds on the operators, as this should be performed by proper global fits \cite{Buckley:2015lku,Hartland:2019bjb,Brivio:2019ius,Biekoetter:2018ypq,Ellis:2018gqa,daSilvaAlmeida:2018iqo,Ellis:2020unq}, but we aim to understand the potential impact of RGE effects.

We use a selection of top observables, specifically: the $t \bar t$ cross-section at $\sqrt{s} = 8$ and $13$ TeV, inclusive and differential in $m_{t \bar t}$, and the top asymmetry $A_C$ at $8$ TeV differential in the $t \bar t$ rapidity and velocity, as described in Table \ref{tab:ToyFitObseravables}. Our fit is performed using the best available SM predictions, as in the last column of the Table,
together with SMEFT predictions at LO QCD and LO EW for a fixed and a dynamical scale, as described in previous Sections. As we selected a set of $t \bar t$ observables that is not particularly sensitive to four-fermion 4H operators, we only constrain four-fermion 2L2H operators, both colour-octets and colour-singlets. 

\begin{table}[h]
	\centering
	\scalebox{0.8}{
	\begin{tabular}{|ccccccc|}\hline
		Experiment & $\sqrt{s}$ [TeV] & $\mathcal{L}$ [fb$^{-1}$] & Channel & Observable & Ref. & SM Th.~Ref. \\ \hline\hline
		ATLAS & 8 & 20.3 & Dilepton & $\sigma_{t\bar{t}}$ & \cite{ATLAS:2014nxi} & {\small NNLO+NNLL QCD, NLO EW}\, \cite{Czakon:2011xx} \\
		CMS & 8 & 19.7 & Lepton+jets & $dA_C/dy_{t\bar{t}}$ [3 bins] & \cite{CMS:2015pob} & {\small NNLO QCD, NLO EW} \, \cite{Czakon:2017lgo}\\
		ATLAS & 8 & 20.3 & Lepton+jets & $dA_C/d\beta_{t\bar{t}}$ [3 bins] & \cite{ATLAS:2015jgj} & {\small NNLO QCD, NLO EW} \cite{Czakon:2017lgo} \\
		CMS & 8 & 19.6 & Lepton+jets & $\sigma_{t\bar{t}}$ & \cite{CMS:2016csa} & {\small NNLO+NNLL QCD, NLO EW}\,\cite{Czakon:2011xx} \\
		CMS & 8 & 19.7 & $e \mu$ & $\sigma_{t\bar{t}}$ & \cite{CMS:2016yys} & {\small NNLO+NNLL QCD, NLO EW}\,\cite{Czakon:2011xx} \\
		ATLAS & 8 & 20.2 & Lepton+jets & $\sigma_{t\bar{t}}$ & \cite{ATLAS:2017wvi} & {\small NNLO+NNLL QCD, NLO EW}\,\cite{Czakon:2011xx} \\
		CMS & 13 & 35.9 & Dilepton & $d\sigma_{t\bar{t}}/dm_{t\bar{t}}$ [7 bins] & \cite{CMS:2018adi} & {\small NNLO+NNLL QCD, NLO EW} \cite{Czakon:2019txp} \\
		ATLAS & 13 & 36 & Lepton+jets & $d\sigma_{t\bar{t}}/dm_{t\bar{t}}$ [7 bins] & \cite{ATLAS:2019hxz} & {\small NNLO+NNLL QCD, NLO EW} \cite{Czakon:2019txp} \\	
		ATLAS & 13 & 139 & Lepton+jets & $\sigma_{t\bar{t}}$ & \cite{ATLAS:2020aln} & {\small NNLO+NNLL QCD, NLO EW}\,\cite{Czakon:2011xx} \\
		CMS & 13 & 137 & Lepton+jets & $\sigma_{t\bar{t}}$ & \cite{CMS:2021vhb} & {\small NNLO+NNLL QCD, NLO EW}\,\cite{Czakon:2011xx} \\ \hline\hline
		\end{tabular}}
	\caption{Top pair production observables considered in the fit, together with the relevant SM theoretical prediction used and their accuracy.}
	\label{tab:ToyFitObseravables}
\end{table}

Constraints are set on the values of the Wilson coefficients at $\mu_0 = 2 \, \text{TeV}$ under three RGE scenarios. The first one, indicated with {\it "No running"}, is obtained setting the anomalous dimension ${\bm \gamma}$ to zero. In this scenario identical constraints are obtained at any scale as the coefficients do not run. This reproduces the past state-of-the art in SMEFT fits, where RGE effects were not included. The second scenario, {\it "Fixed scale"}, amounts to setting $\mu_0$ to 2 TeV and running down to $\mu = m_\text{top}$, and the third scenario {\it "Dynamical scale"} sets $\mu_0$ to 2 TeV and runs to $\mu$, a dynamical quantity evaluated on an event-by-event basis, that we take to be $H_T/2$. Our three scenarios only differ with respect to the EFT scale $\mu$, in all three cases the SM parameters are renormalised at the dynamical scale $\mu_R = \mu_F = H_T/2$.

\begin{figure}[h!]
    \centering
    \includegraphics[width=.85\linewidth]{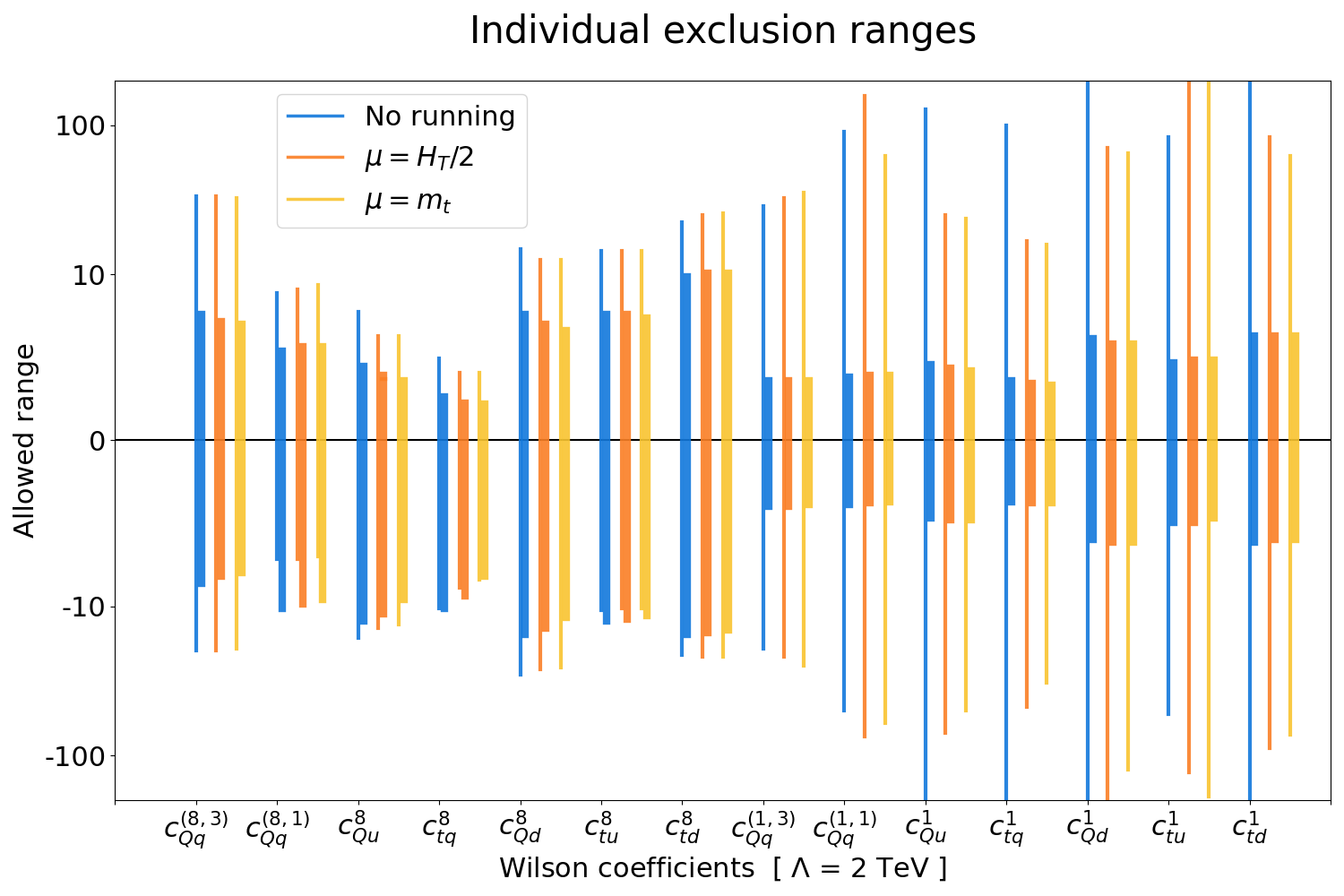}
	\caption{Plot of the results presented in Table \ref{tab:ToyFitResults}. The three RGE cases are distinguished by colour as described by the legend, while the order in $c/\Lambda^2$ is distinguished by line thickness, with the thick line corresponding to the fit using the full $\mathcal O(c^2/\Lambda^4)$ SMEFT predictions and the thin line corresponding to only including the $\mathcal O(c/\Lambda^2)$ component. }
     \label{fig:fit1}
\end{figure}

\begin{figure}[h!]
    \begin{subfigure}{.5\textwidth}
      \centering
      \includegraphics[width=\linewidth]{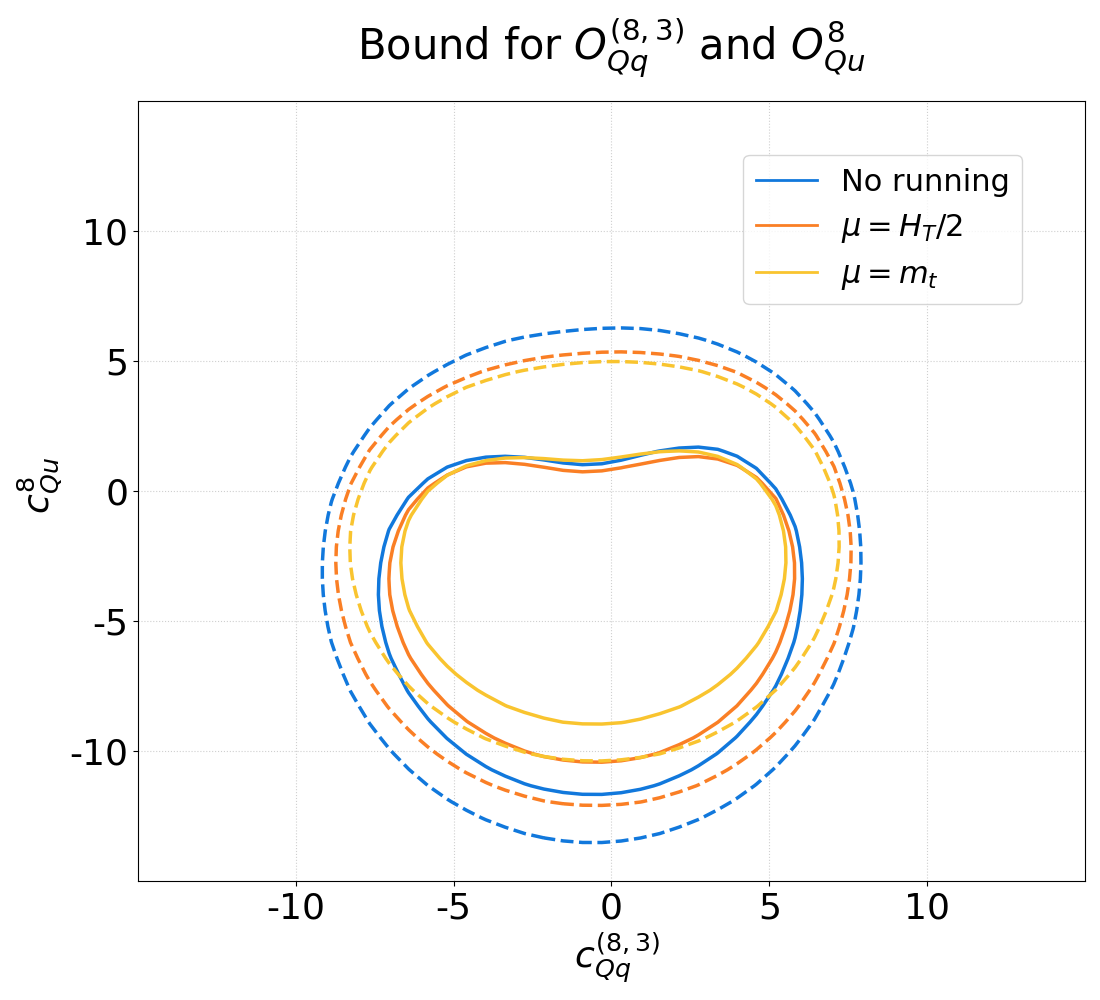}
      \subcaption{}
      \label{fig:fit2_1}
    \end{subfigure}%
    \begin{subfigure}{.5\textwidth}
      \centering
      \includegraphics[width=\linewidth]{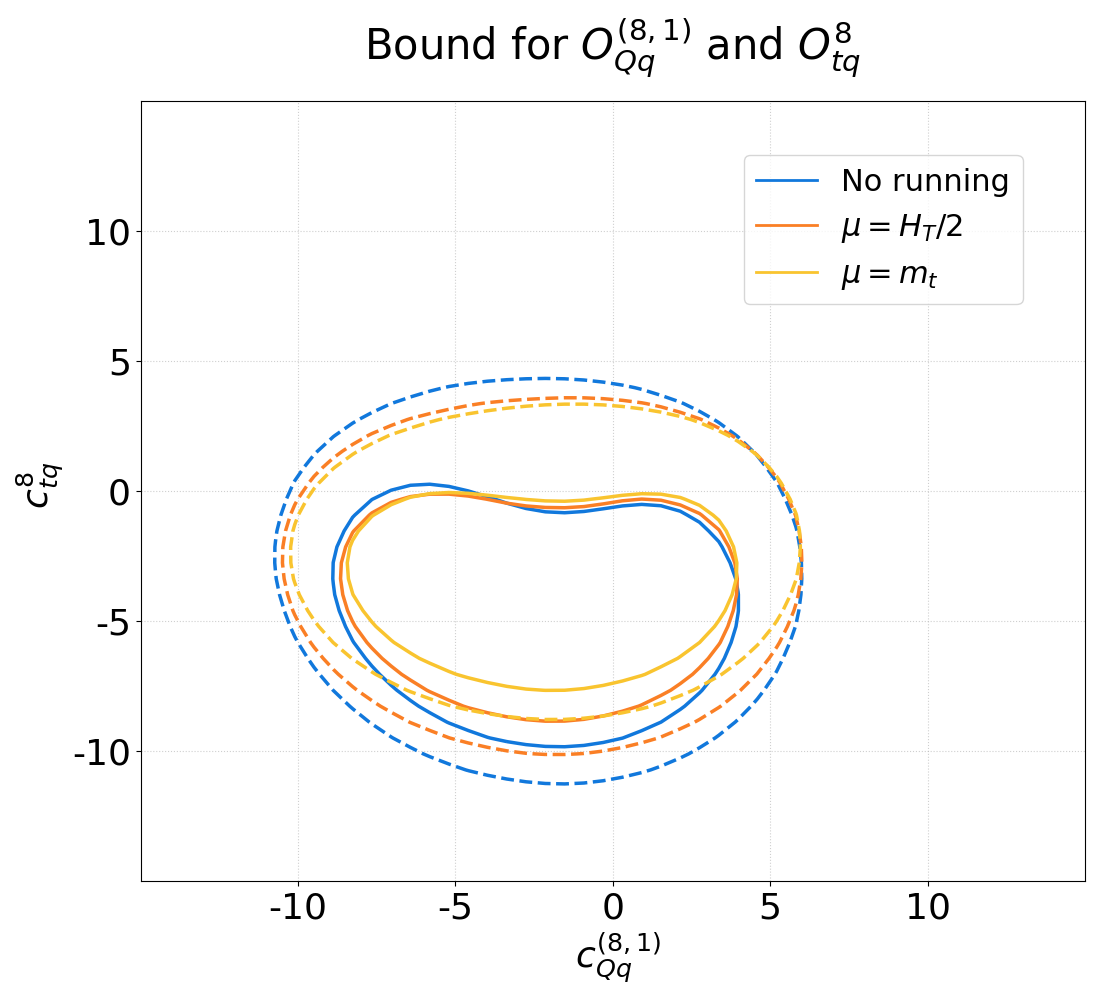}
      \subcaption{}
      \label{fig:fit2_2}
    \end{subfigure}    
    
	\caption{68\% (solid) and 95\% (dashed) allowed regions for the Wilson coefficients $c_{Qu}^{8}$ and $c_{Qq}^{(8,3)}$, left, and $c_{tq}^{8}$ and $c_{Qq}^{(8,1)}$, right, under the three RGE conditions described in the text: no running (blue), dynamical scale (orange), and fixed scale (yellow). Wilson coefficients are evaluated at $\mu_0 = 2 \, \text{TeV}$, the NP scale is also set to $\Lambda = 2 \, \text{TeV}$. Our EFT predictions include both the linear and quadratic term. }
	\label{fig:fit}
\end{figure}

We present the result of our fit in the following Table and Figures. In Table \ref{tab:ToyFitResults}, we show the 2$\sigma$ allowed range for each four-fermion 2L2H Wilson coefficient individually, under the three RGE scenarios described above, for a fit at order $\mathcal O(c/\Lambda^2)$ ("{\it linear}") and $\mathcal O(c/\Lambda^2) + \mathcal O(c^2/\Lambda^2)$  ("{\it quadratic}"). The same results are plotted in Figure \ref{fig:fit1}. In Figures \ref{fig:fit2_1} and \ref{fig:fit2_2} we show the 1 and 2 $\sigma$ contours for selected pairs of Wilson coefficients. 

We first note that given the current sensitivity of experimental measurements, there is a significant difference between the linear and quadratic bounds. For the colour-octets, linear bounds are less stringent by up to factors of a few, whilst colour-singlets remain essentially unconstrained at linear level, before RGE effects are applied. Once the RGE flow is considered, as colour-singlets progressively run into their octet partner, experimental constraints become more and more stringent. With current experimental precision, the bounds on the colour singlet operators remain extremely loose even when RGE effects are considered. At the quadratic level, on the other hand, we find that the individual bounds depend mildly on the scale choice, with the three scenarios differing by at most 10-20\%. 

 In the case of 2-dimensional exclusion regions, we find that, while not dramatically altering the picture, RGE corrections can amount to significant numerical shifts, typically of order of the spread between the 1 and 2 $\sigma$ contours. Currently all Wilson coefficients remain consistent with zero, and RGE effects just shift the allowed intervals. It is worth noting though that as precision of the experimental measurements progressively improves and in the case where a deviation from the SM is confirmed, these effects will be crucial to reliably characterise New Physics. 

 \begin{table}[H]
	\centering
	\scalebox{0.8}{
	\begin{tabular}{|c|ccc|ccc|}\hline
	 \multirow{2}{*}{Wilson Coeff.} & \multicolumn{3}{c|}{Linear fit} &\multicolumn{3}{c|}{Quadratic fit}  \\
	  & No Running & Dynamical Scale & Fixed Scale & No Running & Dynamical Scale & Fixed Scale  \\ \hline
$c_{Qq}^{(8,3)}$ &  $[-19,34]$  &  $[-20,33]$  &  $[-19,33]$  &  $[-8.7,7.5]$  &  $[-8.2,7.2]$  &  $[-7.9,6.9]$  \\
$c_{Qq}^{(8,1)}$ &  $[-7,9]$  &  $[-7,9]$  &  $[-7,9]$  &  $[-10.4,5.5]$  &  $[-10.0,5.7]$  &  $[-9.7,5.7]$  \\
$c_{Qu}^{8}$ &  $[-16,8]$  &  $[-14,6]$  &  $[-13,6]$  &  $[-12.5,4.3]$  &  $[-11.4,3.7]$  &  $[-9.7,3.4]$  \\
$c_{tq}^{8}$ &  $[-11,5]$  &  $[-9,4]$  &  $[-9,4]$  &  $[-10.4,2.6]$  &  $[-9.5,2.3]$  &  $[-8.2,2.1]$  \\
$c_{Qd}^{8}$ &  $[-29,14]$  &  $[-27,13]$  &  $[-26,13]$  &  $[-15.6,7.5]$  &  $[-14.0,6.9]$  &  $[-12.1,6.6]$  \\
$c_{tu}^{8}$ &  $[-11,14]$  &  $[-10,15]$  &  $[-10,15]$  &  $[-12.5,7.6]$  &  $[-12.1,7.6]$  &  $[-11.4,7.5]$  \\
$c_{td}^{8}$ &  $[-22,23]$  &  $[-22,25]$  &  $[-22,26]$  &  $[-15.6,10.0]$  &  $[-15.1,10.2]$  &  $[-14.3,10.2]$  \\
$c_{Qq}^{(1,3)}$ &  $[-19,29]$  &  $[-23,33]$  &  $[-24,35]$  &  $[-4.1,3.6]$  &  $[-4.0,3.6]$  &  $[-3.9,3.6]$  \\
$c_{Qq}^{(1,1)}$ &  $[-49,90]$  &  $[-77,154]$  &  $[-60,59]$  &  $[-3.9,3.8]$  &  $[-3.8,3.8]$  &  $[-3.7,3.8]$  \\
$c_{Qu}^{1}$ &  $[-300,124]$  &  $[-71,25]$  &  $[-50,24]$  &  $[-4.7,4.6]$  &  $[-4.8,4.4]$  &  $[-4.8,4.2]$  \\
$c_{tq}^{1}$ &  $[-207,103]$  &  $[-47,17]$  &  $[-32,17]$  &  $[-3.7,3.6]$  &  $[-3.8,3.4]$  &  $[-3.8,3.3]$  \\
$c_{Qd}^{1}$ &  $[-450,995]$  &  $[-211,70]$  &  $[-126,65]$  &  $[-6.0,6.1]$  &  $[-6.1,5.9]$  &  $[-6.0,5.7]$  \\
$c_{tu}^{1}$ &  $[-52,82]$  &  $[-128,323]$  &  $[-189,196]$  &  $[-5.0,4.8]$  &  $[-4.9,4.8]$  &  $[-4.7,4.8]$  \\
$c_{td}^{1}$ &  $[-268,207]$  &  $[-89,84]$  &  $[-72,64]$  &  $[-6.2,6.3]$  &  $[-6.1,6.3]$  &  $[-5.9,6.3]$  \\
	\hline
		\end{tabular}
		}
	\caption{Numerical results of our fit on the LHC top data in Table \ref{tab:ToyFitObseravables} under the three RGE conditions described in the text, no running, dynamical scale ($\mu = H_T/2$), and fixed scale ($\mu = m_\text{top}$). The reported range for each operator is the 95\% CL interval for the respective Wilson coefficient, in units of $1/(2 \, \text{TeV})^2$, evaluated at $\mu_0 = 2 \, \text{TeV}$. Results in the leftmost three columns are obtained from a fit with linear SMEFT contributions only, while results in the right three columns refer to a fit using the full linear + quadratic SMEFT contributions.}
	\label{tab:ToyFitResults}
\end{table}

\section{Conclusion}
\label{Sec:Conclusions}
We have presented a first study of RGE effects in SMEFT interpretations of collider data. We extracted the anomalous dimension matrix at order $\alpha_s$ for a selection of dimension-6 operators from the counterterms of the \texttt{SMEFT@NLO} model 
and implemented the RGE flow into \texttt{MadGraph5\_aMC@NLO}. This implementation provides a practical way of including RGE effects in actual SMEFT analyses. In particular, our implementation allows, at LO, the computation of RGE running and mixing at run time and on an event-by-event basis, which is necessary to reliably predict differential distributions in the SMEFT. 

To illustrate the setup of our implementation, we focused on top pair production at the LHC, considering five bosonic operators and nineteen four-fermion operators. We studied in detail the impact of including or not including RGE effects, and the impact of employing a fixed or dynamical scale. The running and mixing of Wilson coefficients were shown to shift the theoretical predictions for cross-sections and differential distributions by $\mathcal O(\text{10 - 30\%})$ depending on the operator, with up to $\mathcal O(\text{100 \%})$ deviations observed in special cases. 

We demonstrated the effect of the renormalisation group flow on bounds on Wilson coefficients by performing a toy fit in the top sector. Different scale choices are shown to distort, sometimes significantly, the constraints obtained from experimental data, highlighting the importance of properly including RGE effects in global fits and in SMEFT theoretical predictions in general. 

We note that our toy fit only included $t \bar t$ observables, for most of which the  natural scale is of order $\sim 2 m_{\text{top}}$. When observables from different sectors and more high-energy differential distributions are included in a global fit, we expect that RGE effects will be even more prominent.

We have implemented the RGE evolution to a fixed or dynamical scale in a user-friendly and public setup, available within \texttt{Madgraph5\_aMC@NLO} at version 3.4.0 and beyond. We also released the anomalous dimension matrix at order $\alpha_s$ of the operators we considered, within the \texttt{SMEFTatNLO} model. Extending the implementation to other operators is straightforward. 

Going beyond the SMEFT, the running and mixing implementation can be used for other New Physics scenarios which involve running couplings. Our implementation already steps in this direction, by allowing a more general RGE than the one we considered for the SMEFT. This work paves the way for including running and mixing effects in all future interpretations of LHC measurements. 

\section*{Acknowledgements}
We would like to thank C\'eline Degrande, Gauthier Durieux, Hesham el Faham, Benjamin Fuks, Ken Mimasu, Julie Pagès, Hua-Sheng Shao and Cen Zhang for enlightening discussions. We would also like to thank Aneesh Manohar for helpful and kind discussions regarding the comparison with previous works. We are grateful to Federica Fabbri and the ATLAS Collaboration, and to Marco Zaro, for supplying us with additional results needed for our fit. RA and FM's research was supported by the F.R.S.-FNRS with the EOS - be.h project n. 30820817, the F.R.S-FNRS project no. 40005600 and the FSR Program of UCLouvain. RA, OM and FM acknowledge support by FRS-FNRS (Belgian National Scientific Research Fund) IISN projects 4.4503.16. CS and EV are supported by the European Union’s Horizon 2020 research and innovation programme under the EFT4NP project (grant agreement no. 949451) and by a Royal Society University Research Fellowship through grant URF/R1/201553. Computational resources have been provided by the supercomputing facilities of the Université catholique de Louvain (CIS- M/UCL) and the Consortium des Équipements de Calcul Intensif en Fédération Wallonie Brux- elles (CÉCI) funded by the Fond de la Recherche Scientifique de Belgique (F.R.S.-FNRS) under convention 2.5020.11 and by the Walloon Region.

\clearpage

\appendix

\section{Conventions for operators with repeated currents}
\label{sec:comparison}
In this Appendix we describe the different conventions regarding SMEFT operators with repeated currents, in particular focusing on the comparison between this work and Ref.~\cite{Jenkins:2013zja,Jenkins:2013wua,Alonso:2013hga}.

We consider for concreteness the operator:
\begin{equation}
    Q_{\substack{uu \\ prst}} = (\bar u_p \gamma^\mu u_r) (\bar u_s \gamma_\mu u_t).
\end{equation} 
Due to the repeated $\bar u \gamma u$ bilinear, this operator enjoys the symmetry:
\begin{equation}
    Q_{\substack{uu \\ prst}} = Q_{\substack{uu \\ stpr}}. \label{symm_uu}
\end{equation} 
In our convention, if $pr \neq st$, the Lagrangian contains {\it one} between $Q_{\substack{uu \\ prst}}$ and $Q_{\substack{uu \\ stpr}}$:
\begin{equation}
    \mathcal L_1 \ \supset \ \sum_{pr=st} C_{\substack{uu \\ prpr}} Q_{\substack{uu \\ prpr}} +  \sum_{\substack{pr \neq st \\ \text{unique}}} C_{\substack{uu \\ prst}} \, Q_{\substack{uu \\ prst}}, \label{one}
\end{equation}
where by {\it unique} we mean that only one between e.g. 1133 and 3311 is included in the sum.

In the convention of \cite{Jenkins:2013zja,Jenkins:2013wua,Alonso:2013hga}, the Lagrangian is summed over {\it all} flavor indices regardless of symmetries such as \eqref{symm_uu}. The redundant degrees of freedom $C_{\substack{uu \\ prst}}$ and $C_{\substack{uu \\ stpr}}$ can be broken into a symmetric and anti-symmetric part. Since upon summing over $prst$ and enforcing \eqref{symm_uu} the anti-symmetric part drops out of the Lagrangian, one can replace $C_{\substack{uu \\ prst}} \to 1/2(C_{\substack{uu \\ prst}} + C_{\substack{uu \\ stpr}})$ and $C_{\substack{uu \\ stpr}} \to 1/2(C_{\substack{uu \\ prst}} + C_{\substack{uu \\ stpr}})$, and obtain:
\begin{equation}
    \mathcal L_2 \ \supset \ \sum_{pr=st} C_{\substack{uu \\ prpr}} Q_{\substack{uu \\ prpr}} +  \sum_{\substack{pr \neq st \\ \text{unique}}} 2 \, C_{\substack{uu \\ prst}} \, Q_{\substack{uu \\ prst}}. \label{two}
\end{equation}
We note that while $\mathcal L_1$ and $\mathcal L_2$ are ultimately equivalent, their relative RGE's will be different, especially in the presence of further flavor assumptions, such as our \eqref{eq:conv1}-\eqref{eq:conv2}. 

\section{Generation details}
\label{sec:Gen_details}

This Appendix contains details of our implementation of the RGE in \texttt{Madgraph5\_aMC@NLO} (available since version 3.4.0). As the two-loop accurate RGE is currently not fully known, our implementation is limited to LO event generation only, to preserve the formal accuracy of the LO/NLO perturbative orders. (Therefore \texttt{Madgraph5\_aMC@NLO} out-of-the-box will be able to produce all the results in this paper except for the "NLO" curve in Figure \ref{fig:NLOvsLOcomparison}.)

The new functionalities are available after executing the command:
\begin{verbatim}
install RunningCoupling
\end{verbatim}
This installs the relevant material into \texttt{Madgraph5\_aMC@NLO}, in the form of a {\tt C/Fortran} library \cite{cormen01introduction, NIJENHUIS197899,r8lib}.

The RGE is implemented in the \texttt{running.py} file, that is part of UFO models. This file contains the anomalous dimension matrix in the form of \texttt{Running} Python objects, instances of the class \texttt{Running} defined in \texttt{object\_library.py}, one for each non-zero entry. \smallskip

The structure of \texttt{running.py} is best illustrated by an example. The RGE for $dc_{Qq}^{1,3}/d\log \mu$ contains a term proportional to $c_{Qq}^{8,3}$, with coefficient $(8/3)(\alpha_s/4\pi)$. This is encoded in the following object:
\begin{verbatim}
RGE_7_9 = Running(name        = 'RGE_79',
                run_objects = [[P.cQq13, P.cQq83, P.aS]],
                value       = '(8./3.)/(4.*cmath.pi)')
\end{verbatim}
The string under \texttt{value} can be any valid Python numerical expression; it is evaluated at process generation time, multiplied by $\alpha_s$, and placed in the $c_{Qq}^{(1,3)}-c_{Qq}^{(8,3)}$ position of the anomalous dimension matrix. Once the matrix is reconstructed, with entries not supplied assumed to be zero, the RGE is solved using \eqref{eq:rge_montecarlo}, and the 2-loop accurate expression for $\alpha_s(\mu)$. \smallskip

In this work we have focused on \eqref{eq:rge}, however the current code handles a more general RGE, of the following type:
\begin{equation}
\frac{dc_i(\mu)}{d\log\mu} = \left( \frac{g_s(\mu)}{16 \pi^2} \, v_{ij}^{ \text{\tiny \rm QCD,1}} +  \frac{\alpha_s(\mu)}{4\pi} \, \gamma_{ij}^{ \text{\tiny \rm QCD,1}} \right) \, c_j(\mu).
\label{eq:rge1}
\end{equation} 

From \texttt{Madgraph5\_aMC@NLO} version 3.5.0, the code supports both $v_{ij}^{ \text{\tiny \rm QCD,1}}$ and $\gamma_{ij}^{ \text{\tiny \rm QCD,1}}$ to be arbitrary functions of the model input parameters, such as $g, g', v$. This is needed, e.g., for our $\bm{\gamma_{\text{0/2F}}^{\text{\tiny \rm QCD,1}}}$ in Eq.~\eqref{eq:Gamma0and2F}.

\smallskip 

A new section has been added to the LO Run Card:

\begin{verbatim}
#***********************************************************************
# CONTROL The extra running scale (not QCD)                            *
#    Such running is NOT include in systematics computation            *
#***********************************************************************
True	= fixed_extra_scale ! False means dynamical scale 
172.5	= mue_ref_fixed ! scale to use if fixed scale mode
1.0	= mue_over_ref ! ratio to mur if dynamical scale
\end{verbatim}

If \texttt{fixed\_extra\_scale} is True, the scale $\mu$ is taken to be the constant \texttt{mue\_ref\_fixed}; if \texttt{fixed\_extra\_scale} is False, the scale $\mu$ becomes dynamical and is taken to be \texttt{mue\_over\_ref} times $\mu_R$. For simplicity, we have implemented this scenario by allowing $\mu$ to be a fixed multiple of $\mu_R$, the SM renormalization scale. Therefore, to obtain a dynamical $\mu$ one needs to select a dynamical value for $\mu_R$, and choose the desired, constant, ratio  $\mu/\mu_R$. We note that \texttt{Madgraph5\_aMC@NLO} already allows $\mu_R$ to be an arbitrary function of momenta, therefore the same now applies to the SMEFT scale $\mu$. The case of a fixed $\mu_R$ and dynamical $\mu$ is currently not implemented, but it has little physical motivation. \smallskip

To specify the scale $\mu_0$ in {\tt param\_card.dat}, we follow the SLHA convention \cite{Skands:2003cj} which allows to associate an input scale to any block of the Parameter Card. Following is an example, in the context of the {\tt SMEFTatNLO} model:
\begin{verbatim}
###################################
## INFORMATION FOR DIM62F
###################################
Block dim62f Q= 2000
    1 1.000000e+00 # cpl1
    2 0.000000e+00 # cpl2
    3 0.000000e+00 # cpl3
\end{verbatim}
The value after {\tt Q=} is the scale in GeV, $\mu_0$ in the above text, at which the couplings are provided. {\tt MadGraph5\_aMC@NLO} automatically checks that all parameters mixing into each other are provided at the same scale, and will stop the computation if this is not the case. On the other hand, if the anomalous dimension matrix is block diagonal, then the code allows each block to have its own $\mu_0$.

\section{Additional results for differential distributions}
\label{sec:Additional_differential}
In this Section, we provide plots similar to Figure \ref{fig:mtt_scales}, for additional selected 2L2H and 4H four-fermion Wilson coefficients entering $t \bar t$ production.

\begin{figure}[!htb]
	\centering
	\includegraphics[width=.46\textwidth]{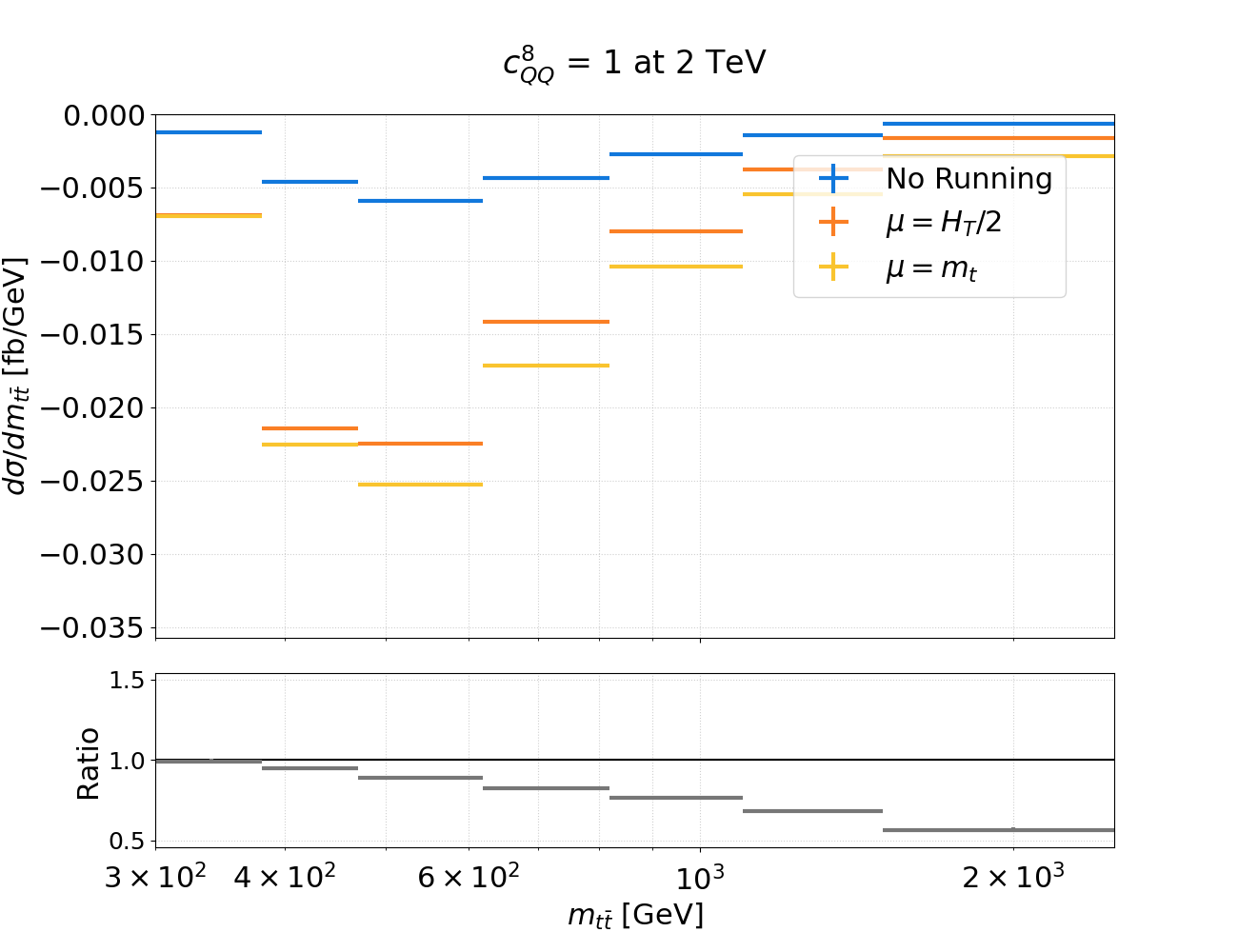}
	\includegraphics[width=.46\textwidth]{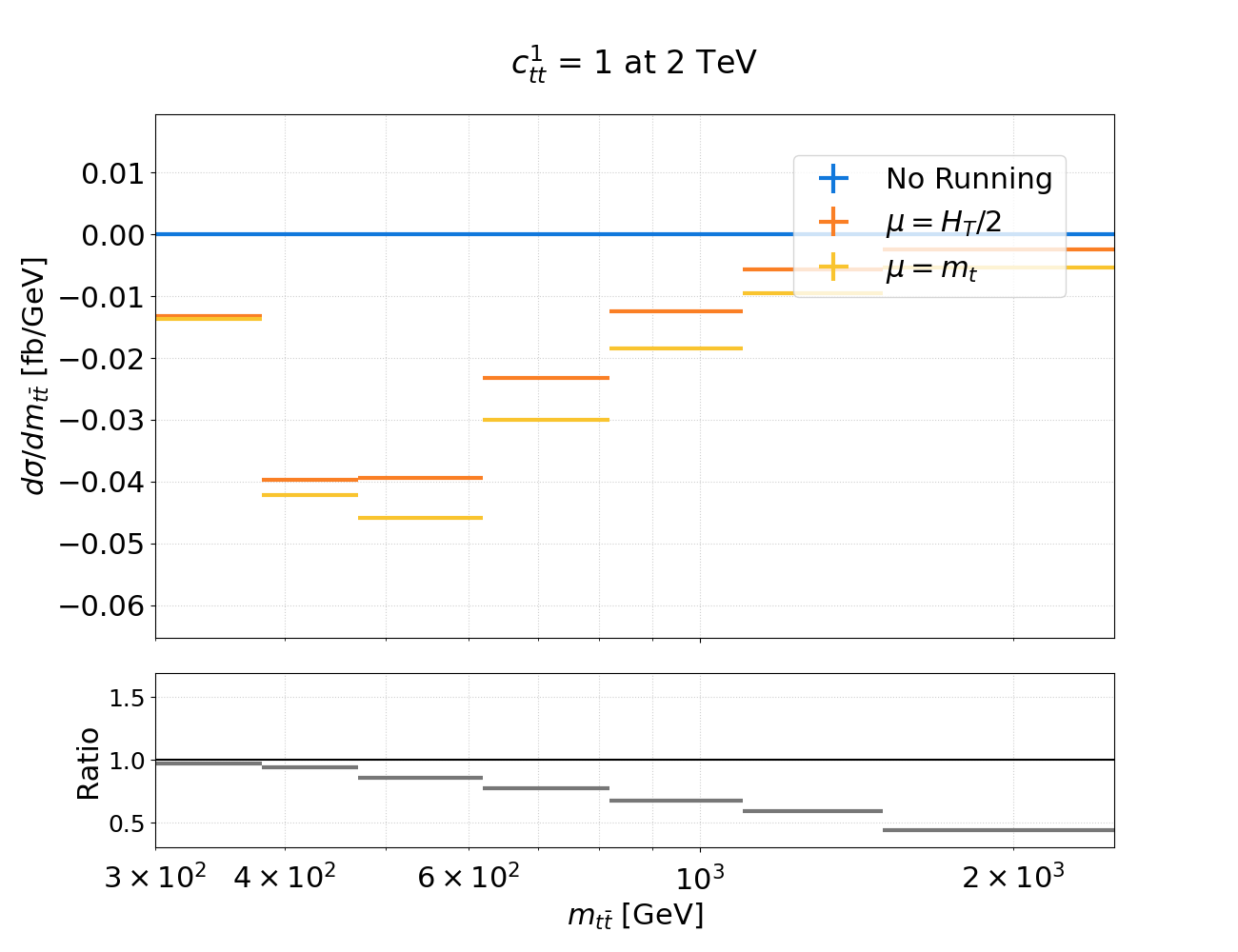}
	\caption{Same as Fig.~\ref{fig:mtt_scales} for the 4H operators $\mathcal O_{QQ}^{8}$ and $\mathcal O_{tt}^1$.}
\end{figure}

\begin{figure}[!htb]
	\centering
	\includegraphics[width=.46\textwidth]{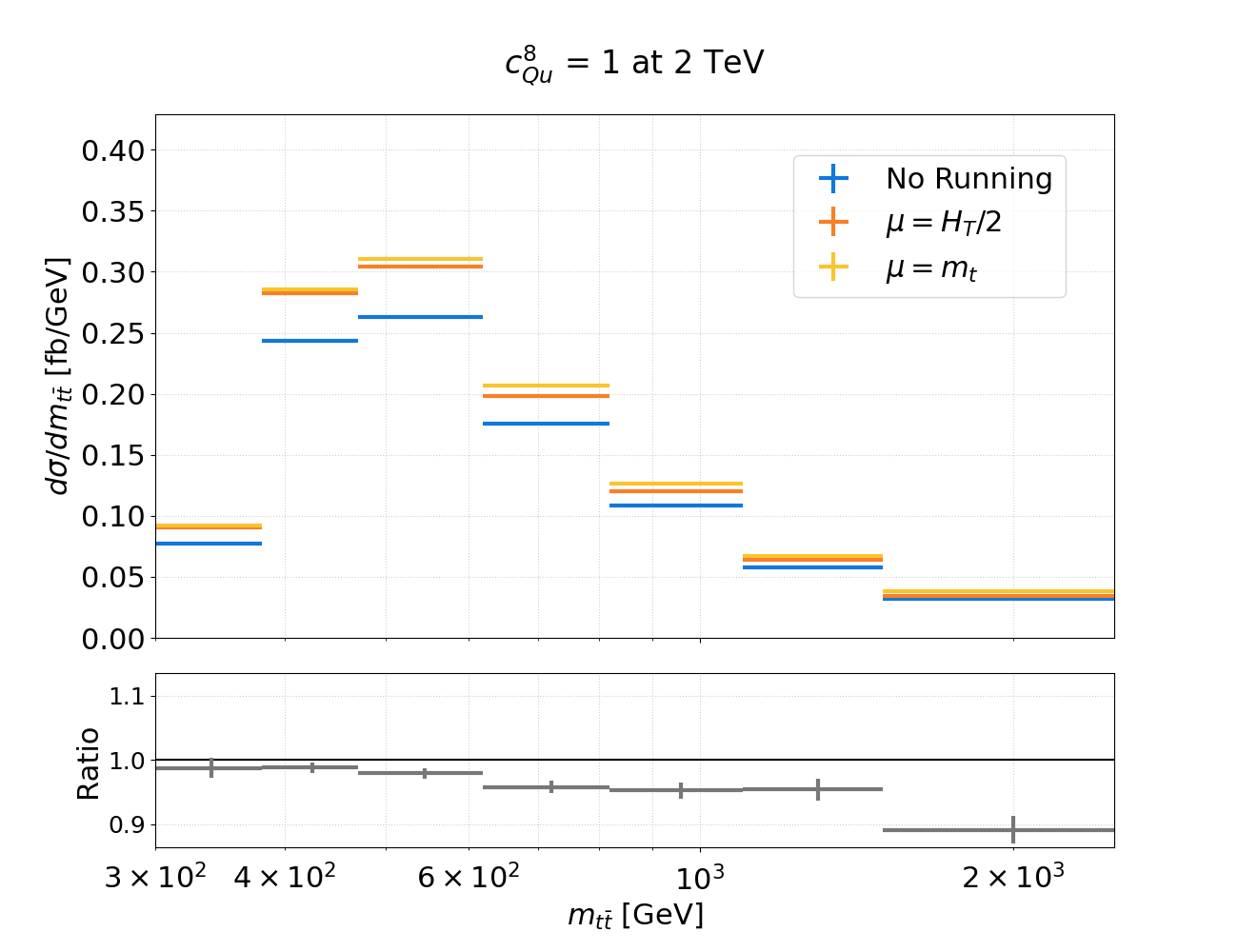}
	\includegraphics[width=.46\textwidth]{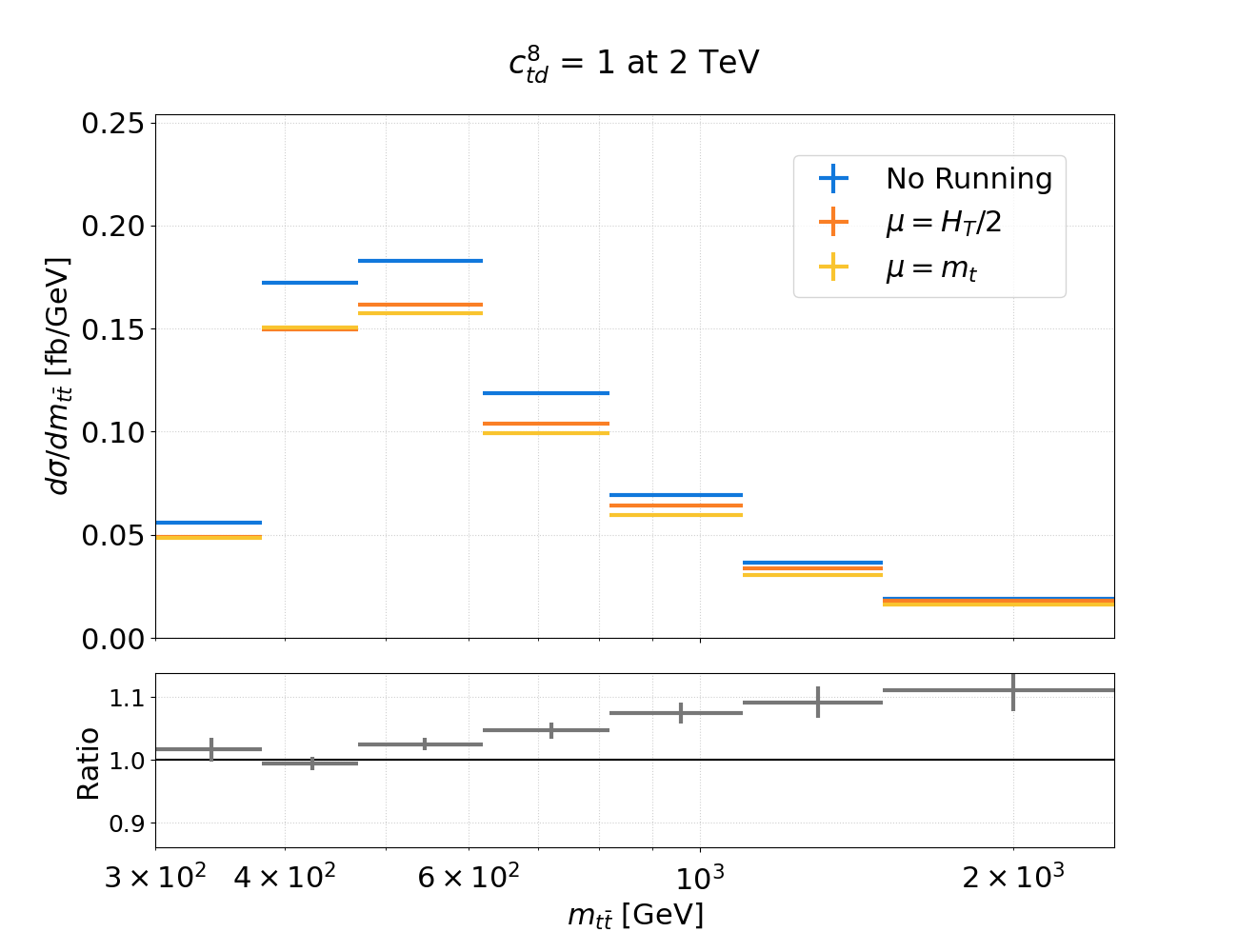}
	\includegraphics[width=.46\textwidth]{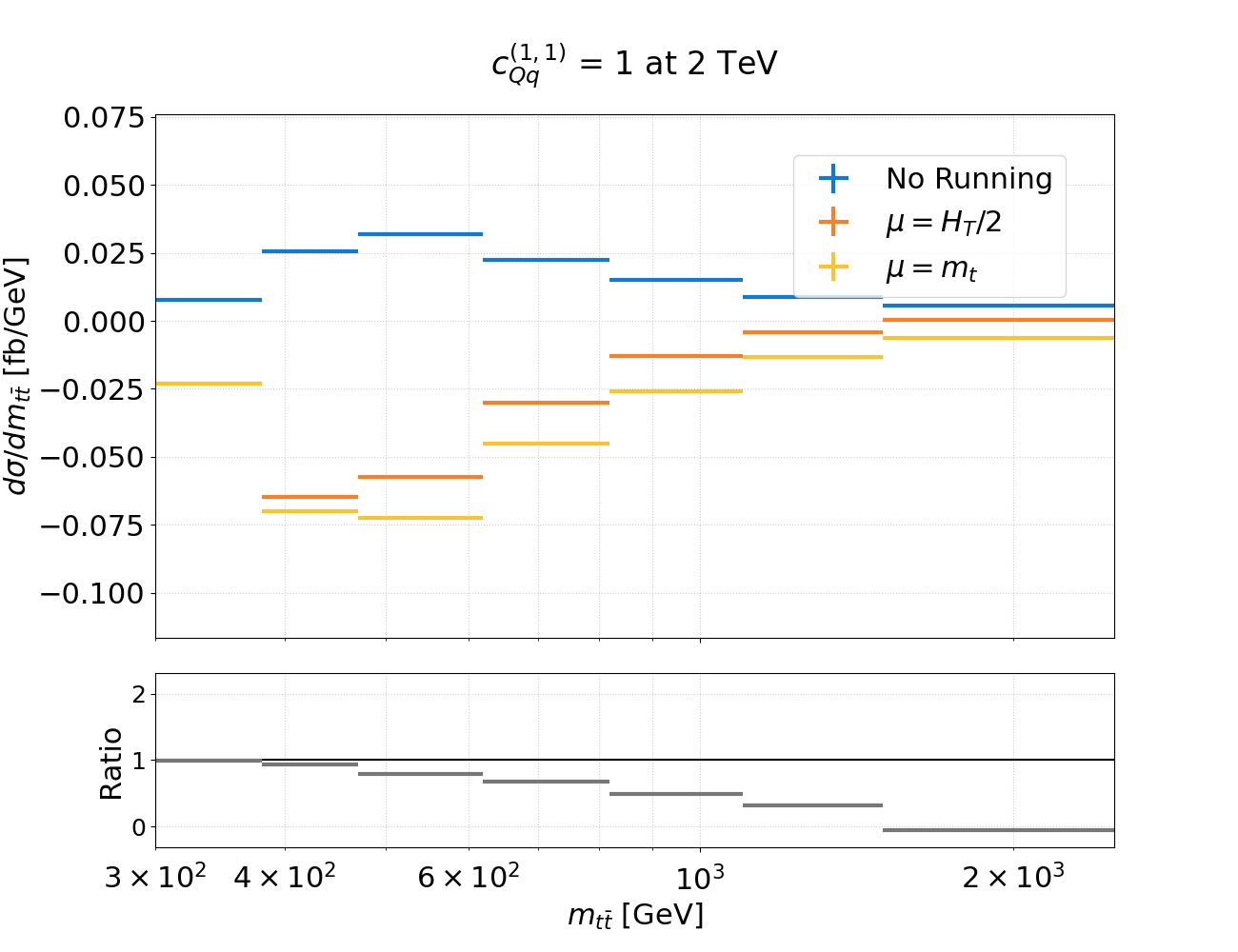}
	\includegraphics[width=.46\textwidth]{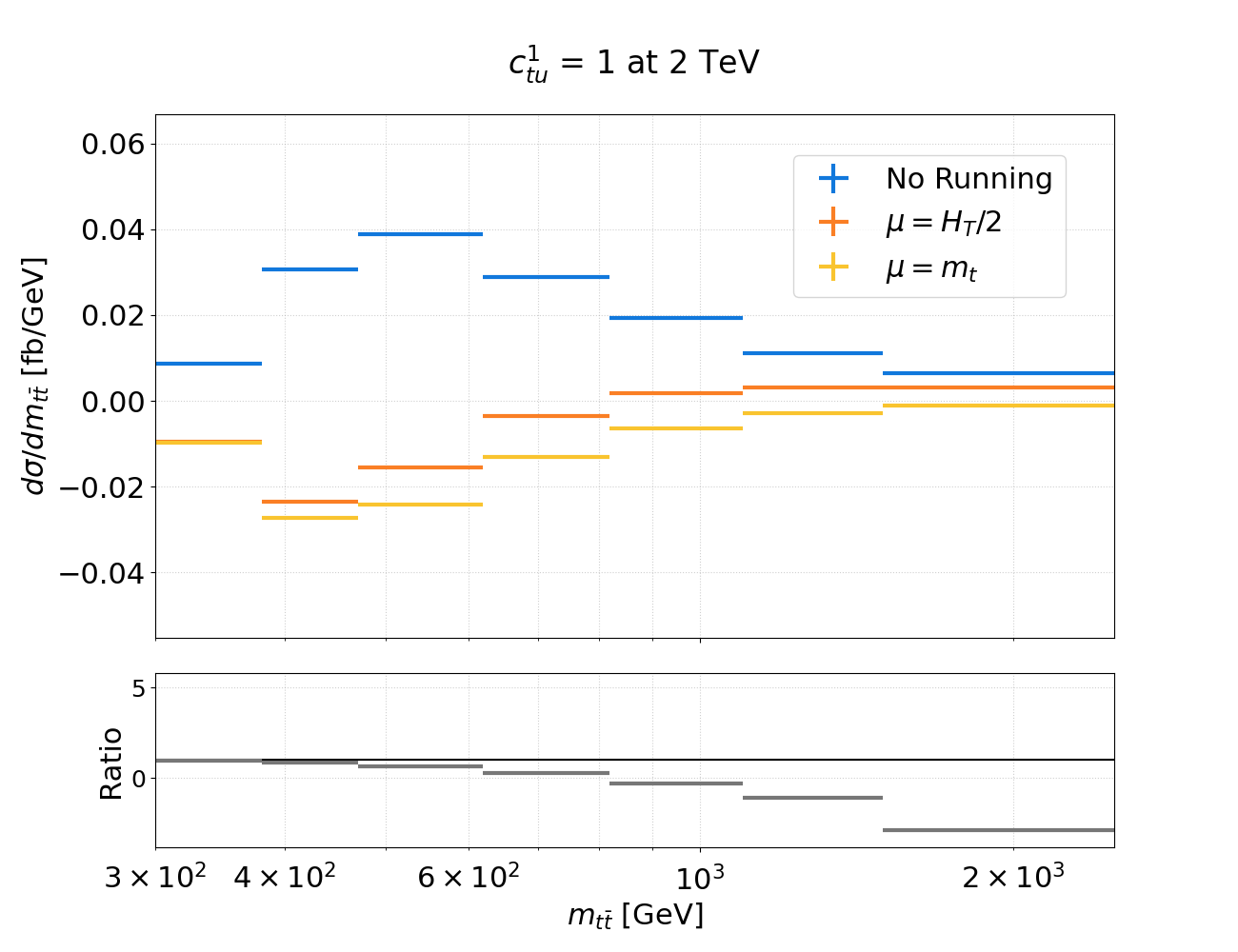}
	\caption{Same as Fig.~\ref{fig:mtt_scales} for the 2H2L colour-singlet operators $\mathcal O_{Qu}^8$ and $\mathcal O_{td}^8$, top, and $\mathcal O_{Qq}^{(1,1)}$ and $\mathcal O_{tu}^1$, bottom.}
\end{figure}

\bibliographystyle{JHEP}
\bibliography{references}

\end{document}